\documentclass[fleqn,usenatbib]{mnras}

\usepackage{newtxtext,newtxmath}

\usepackage[T1]{fontenc}

\DeclareRobustCommand{\VAN}[3]{#2}
\let\VANthebibliography\thebibliography
\def\thebibliography{\DeclareRobustCommand{\VAN}[3]{##3}\VANthebibliography}

\usepackage{graphicx}	
\usepackage{amsmath}	

\usepackage{hyperref}
\hypersetup{
    colorlinks=true,
    linkcolor=cyan,
    citecolor=cyan,
    urlcolor=cyan,
    }

\usepackage{orcidlink}

\usepackage{placeins}

\title[Symbiotic binaries in the \textit{Gaia} data I.]{Symbiotic binaries in the \textit{Gaia} data. I. Known symbiotics in DR3 and FPR
}

\author[J. Merc]{
J. Merc$^{1,2}$\thanks{E-mail: jaroslav.merc@mff.cuni.cz}\orcidlink{0000-0001-6355-2468}
\\
$^{1}$Astronomical Institute, Faculty of Mathematics and Physics, Charles University, V Hole\v{s}ovi\v{c}k{\'a}ch 2, 180 00 Prague, Czech Republic\\
$^{2}$Instituto de Astrof\'isica de Canarias, Calle Vía Láctea, s/n, E-38205 La Laguna, Tenerife, Spain
}

\date{Accepted 2026 January 16. Received 2026 January 16; in original form 2025 September 8}

\pubyear{\the\year{}}

\begin{document}
\label{firstpage}
\pagerange{\pageref{firstpage}--\pageref{lastpage}}
\maketitle

\begin{abstract}
Symbiotic stars are long-period interacting binaries composed of an evolved giant and a hot compact companion. Their complex spectra and variability make them both astrophysically valuable and observationally challenging. We investigate how known symbiotic stars are represented in \textit{Gaia} DR3 and the Focused Product Release (FPR), assess the reliability of these data, and evaluate their usefulness for candidate analysis and searches for new systems. We crossmatched \textit{Gaia} DR3 and the FPR with confirmed symbiotic stars from the New Online Database of Symbiotic Variables and examined their astrometric, photometric, and spectroscopic data, along with derived products such as astrophysical parameters, orbital solutions, variability properties, and emission-line classifications. Astrometric data reliably constrain the position of symbiotics in the color-magnitude diagram, aiding searches for new systems, while RUWE is generally not a reliable indicator of their binarity. Most symbiotics are variable in \textit{Gaia} photometry. Mean and epoch radial velocities, as well as inferred orbital solutions, are broadly consistent with the literature, and we provide the first tentative orbital solutions for two systems. Effective temperatures and metallicities are unreliable due to contamination from nebular continuum and strong emission lines. H$\alpha$ emission is detected in nearly all symbiotics, making it a robust diagnostic. Additionally, resolved companions were identified for two systems, and one previously confirmed symbiotic star was reclassified as non-symbiotic. \textit{Gaia} DR3 provides a rich dataset for the study of symbiotic binaries. The forthcoming DR4 promises a major leap forward with its longer time baseline, new data products, and epoch data.
\end{abstract}

\begin{keywords}
binaries: symbiotic --- variables: general --- Stars: oscillations --- Stars: emission-line, Be --- Catalogs --- Surveys
\end{keywords}



\section{Introduction}
The primary objective of the European Space Agency’s \textit{Gaia} mission is to deliver unprecedented astrometric data for billions of stars \citep{2016A&A...595A...1G}. However, the mission has produced a~wealth of additional data products that extend far beyond astrometry. The third data release (\textit{Gaia} DR3; \citealt{2023A&A...674A...1G}), issued in June 2022, based on 34 months of data (July 25, 2014 -- May 28, 2017), has transformed virtually every field of astrophysics.

In this work, we turn to symbiotic stars, wide interacting binaries composed of cool red giant branch (RGB) or asymptotic giant branch (AGB) donors and hot compact accretors, most often white dwarfs (see reviews by \citealt{2012BaltA..21....5M}; \citealt{2019arXiv190901389M}; \citealt{2025Galax..13...49M}). This paper is the first in a planned series exploring symbiotic stars with \textit{Gaia} data. Here we concentrate on confirmed systems listed in the New Online Database of Symbiotic Variables (version from February 3, 2024)\footnote{We also include new symbiotic stars discovered recently during our search with Southern African Large Telescope (\citealt{2026MNRAS.545S2146M}) and unpublished discoveries from our ongoing project (\citealt{Merc+Gaia3}, in prep.).} \citep[NODSV;][]{2019RNAAS...3...28M,2019AN....340..598M,Merc+NODSV2025}, while the discussion of literature candidates and systematic searches for additional symbiotics in DR3, motivated by the current work, will be presented in subsequent papers.

The paper is organized as follows. In Section~\ref{sec:sample}, we describe the sample of confirmed symbiotic stars and briefly summarize the types of data available for them in \textit{Gaia}~DR3 and Focused Product Release (FPR). In Section~\ref{sec:astrometry}, we present the astrometric information on symbiotic stars, including associated parameters such as RUWE and their applicability in searches for symbiotic systems. We also discuss distances and the $z$ distribution. Section~\ref{sec:photometry} discusses the photometric data, variability classifications, and comparisons between \textit{Gaia} measurements and known variability parameters. We also position symbiotic stars in \textit{Gaia} color-magnitude diagram. In Section~\ref{sec:spectroscopy}, we focus on the \textit{Gaia} $BP/RP$ and RVS spectroscopy of symbiotic stars, emission-line classifications, radial velocity measurements, and orbital properties reported in the Non-Single Star tables. We also compare astrophysical parameters derived as part of \textit{Gaia} DR3 with values obtained in detailed studies of individual objects, and discuss emission-line nature and classification of symbiotic systems. Finally, in Section~\ref{sec:conclusions}, we summarize our results, discuss which \textit{Gaia} DR3 parameters are reliable for the symbiotic-star population, and outline possible ways in which \textit{Gaia} data can be used to search for new members of this class.

\section{Sample of symbiotic stars in \textit{Gaia} DR3} \label{sec:sample}

Our starting sample contained 400 objects\footnote{The full list, together with the corresponding \textit{Gaia} identifiers, is provided as supplementary material to this work.}, of which 71 are located in external galaxies (Draco Dwarf, IC 10, LMC, M31, M33, NGC 185, NGC 205, NGC 6822, and SMC), with the remainder belonging to the Milky Way. We find that 36 sources have no counterpart in the main \textit{Gaia} DR3 catalog ({\tt gaia\_source} table); all but one of these are extragalactic and very faint. The only Galactic source without a DR3 counterpart is the symbiotic X-ray binary 2XMM~J174445.4$-$295046, which is extremely faint in the optical due to heavy reddening.

For the extragalactic systems, those located outside the Draco Dwarf, LMC, and SMC are generally too faint for \textit{Gaia} to provide useful information (their median $G$-band magnitude is 20.5 mag). We therefore omit them from most of the further analysis and retain only the Galactic symbiotics (328 systems), together with ten in the LMC (median $G = 15.5$ mag), twelve in the SMC (median $G = 15.7$ mag), and one in the Draco Dwarf ($G = 16.6$ mag). The analyzed sample thus consists of 351 confirmed symbiotic stars. When relevant, we distinguish between Galactic and extragalactic sources in the following sections.

The crossmatch with \textit{Gaia} DR3 shows that nearly all Galactic sources, with the exception of four, have a full astrometric solution including position, parallax, and proper motion, while the remaining four are limited to a two-parameter solution (position only). Mean photometry is available for every source, although two objects are covered only in the $RP$ band and one only in the $G$ band. Epoch photometry is provided for 331 stars \citep[][]{2023A&A...674A..13E}. $BP/RP$ spectra are available in sampled form for 260 stars and in continuous representation for an additional 62 objects \citep[][]{2023A&A...674A...2D,2023A&A...674A...3M}. Median RVS spectra were published for 13 stars, and median radial velocities are available for 137 objects \citep[][]{2023A&A...674A...5K}. Epoch radial velocities are not included in DR3 for any symbiotic stars, but 9 systems have radial velocities reported in the \textit{Gaia} FPR of October 2023, which targeted long-period variables \citep[][]{2023A&A...680A..36G}.

A few additional catalog matches, some of them discussed later in the text, are noteworthy. Sanduleak’s star and H 1-36 appear in the quasar (QSO) candidates table ({\tt qso\_candidates}), but both with negligible probability of being QSO; ten objects are listed in the Non-Single Star tables ({\tt nss\_two\_body\_orbit;} \citealt{2025A&A...693A.124G}); and EG And is part of the \textit{Gaia} Andromeda Photometric Survey \citep[GAPS;][]{2023A&A...674A...4E}.

\begin{figure*}
\centering
\includegraphics[width=\textwidth]{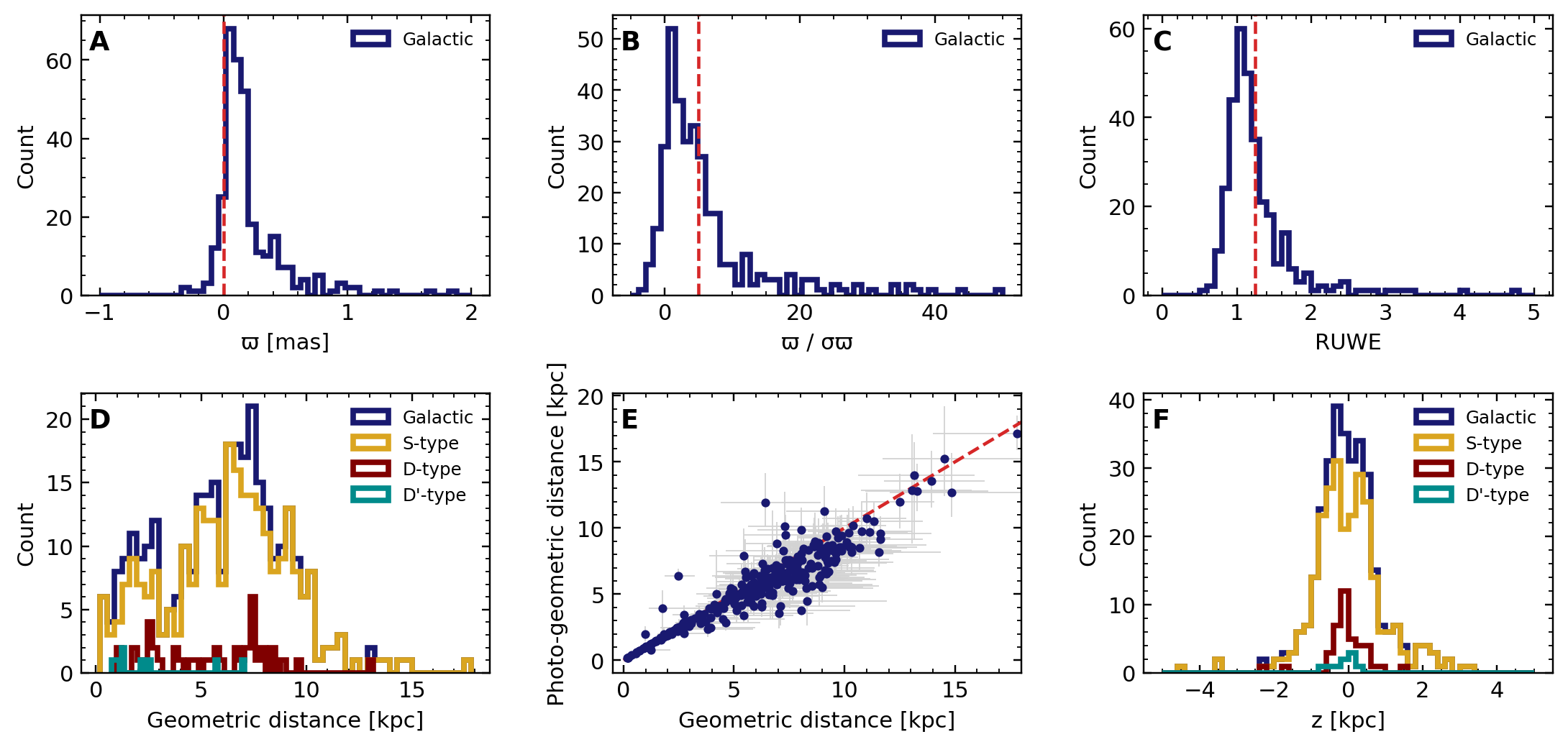}
\caption{Astrometric properties of known symbiotic stars.
\textbf{A:} Distribution of parallaxes from DR3. The red dashed line marks $\varpi = 0$ mas. Seven stars with $\varpi > 2$ mas are not shown.
\textbf{B:} Parallax signal-to-noise ratio. The red dashed line marks $\varpi/\sigma_\varpi = 5$. Three objects with S/N $> 50$ are not shown, including SWIFT J171951.7-300206 (discussed in the text; $\varpi/\sigma_\varpi = 255$).
\textbf{C:} RUWE values. The red dashed line marks RUWE = 1.25. Six objects with RUWE $> 5$ are not shown.
\textbf{D:} Distribution of geometric distances from \citet{2021AJ....161..147B}, color-coded by IR type (S, D, D').
\textbf{E:}~Comparison of geometric and photogeometric distances \citep{2021AJ....161..147B}. The red dashed line indicates the 1:1 relation.
\textbf{F:} Distribution of vertical distances from the Galactic plane, with colors denoting IR types.}
\label{fig:par}
\end{figure*}

\section{Astrometric data}\label{sec:astrometry}

In the discussion of astrometric parameters, we restrict ourselves to Galactic objects. Four stars have no parallax measurement in \textit{Gaia} DR3, 39 have negative parallaxes, and an additional 165 have a~parallax signal-to-noise ratio below 5, leaving 120 systems with a~reasonably measured parallax (expectedly mostly the closer ones). The goodness-of-fit (GOF) statistics yield a seemingly comparable figure, with 155 sources having GOF < 3, which is the threshold for a good fit according to the \textit{Gaia} DR3 documentation. However, 22 of these objects have negative parallaxes, and 76 have parallaxes with signal-to-noise ratios below~5.

In addition, 114 stars from the sample have Renormalized Unit Weight Error (RUWE)\footnote{RUWE contributes to the computation of the GOF. Consequently, when RUWE indicates a poor solution, the GOF is likely to do so as well.} values exceeding 1.25. Values close to 1.0 are expected for sources well described by the single-star astrometric model, while significantly higher values may indicate problems with the solution or the presence of binarity \citep[see, e.g.,][]{2022MNRAS.513.2437P,2024A&A...688A...1C}. Applying the strictest quality cuts, requiring simultaneously a parallax S/N > 5, GOF < 3, and \mbox{RUWE < 1.25}, reduces the sample to only 57 stars out of the initial 328. The distribution of parallaxes and S/N of parallaxes is shown in Figs.~\ref{fig:par}A and~\ref{fig:par}B.

An interesting outcome of this exercise is that only about one-third of known Galactic symbiotic stars have RUWE values above the commonly adopted threshold of 1.25 (Fig.~\ref{fig:par}C). This threshold is specific to a given data release; for DR2, higher values of 1.3–1.4 were typically used, while a value of 1.25 was proposed for DR3 \citep{2022MNRAS.513.2437P}. Above this level, the single-star model does not adequately describe the astrometry, and RUWE is therefore considered an empirical indicator of possible binarity. The threshold itself, however, is sky-position dependent because of the \textit{Gaia} scanning law \citep[see Fig.~3 in][]{2024A&A...688A...1C}. To account for this, we used the {\tt gaiaunlimited} Python package \citep{2023A&A...677A..37C} to obtain position-specific thresholds for individual symbiotic stars and checked whether this changes the inferred "binary fraction". With position-specific thresholds, 109 objects remain above the limit. Compared to the fixed 1.25 cut, two objects are added (with RUWE slightly below 1.25 but above their local threshold), while seven are removed (with RUWE above 1.25 but below their higher local threshold, which in our sample ranges from 1.17 to 1.36). Overall, however, the conclusion remains unchanged: RUWE is not a reliable diagnostic of binarity for this class of interacting binaries and should be applied with caution in searches for new symbiotic stars. We also note that this conclusion is robust across the full range of orbital periods and distances sampled. Neither short- nor long-period systems, nor nearby versus distant symbiotic stars, exhibit any systematic RUWE behaviour that would make this parameter useful for identifying binary properties in these systems.

\subsection{Distances and $z$ distribution}

An important parameter that can be inferred from \textit{Gaia} astrometry is the distance to symbiotic stars. At present, distances are generally very poorly constrained, often estimated only indirectly by assuming absolute magnitudes or other proxies \citep[see discussion in][]{2025A&A...695A..61M}. Consequently, distance-dependent parameters, most notably luminosities, are also uncertain, making direct comparison with evolutionary models difficult. In addition to some parallaxes of symbiotic stars in DR3 being negative or measured with very low S/N, one must also keep in mind that the parallaxes are derived under the assumption of a single star, whereas symbiotics are binaries. The impact of binarity on the astrometric solution depends on the sky position (through the \textit{Gaia} scanning law), but also on the orbital period, mass ratio, and luminosity ratio, which determine the offset between the motion of the center of mass and the center of light \citep[][]{2020MNRAS.495..321P}. Our analysis in \citet{2025A&A...695A..61M} showed that parallaxes of symbiotic binaries can easily be biased by up to a~factor of $\sim$2.

The distribution of distances derived from \textit{Gaia} EDR3 astrometry by \citet{2021AJ....161..147B}, using a probabilistic approach with a~three-dimensional Galactic prior, is shown in Fig.\ref{fig:par}D. It confirms that a large fraction of symbiotic stars lie at distances of several kpc. A~comparison of geometric and photo-geometric distances, which also incorporate colors and apparent magnitudes, motivated by the fact that stars typically occupy restricted regions in the color-magnitude diagram, is also presented (Fig.~\ref{fig:par}E), showing good overall correlation, though the photo-geometric distances are systematically slightly lower. In Fig. \ref{fig:dist}, we compare the geometric distances from \citet{2021AJ....161..147B} with those obtained by simple parallax inversion. The two estimates agree well up to $\sim$5--7 kpc, particularly for sources with relatively high parallax S/N. At larger distances, however, direct inversion of very small parallax values tends to overestimate distances, often placing stars beyond the plausible extent of the Milky Way.

In Fig.~\ref{fig:par}F, we show the vertical distance from the Galactic plane, $z$, derived using the geometric distances of \citet{2021AJ....161..147B}, separating the different infrared (IR) types of symbiotic stars \citep["stellar" or S-type and "dusty" D- and D'-types; see][]{2025Galax..13...49M}. The overall distribution is consistent with the results of \citet{2019ApJS..240...21A} based on \textit{Gaia} DR2, namely a concentration of D-type systems close to the Galactic plane, whereas S-type symbiotics extend to larger $z$. Interestingly, the distribution of S-type stars appears bimodal, with a slight deficit around $z \sim 0$.

\subsection{Resolved companions}

Finally, we also used \textit{Gaia} DR3 astrometry to search for resolved companions to symbiotic stars. The expected fraction of such systems is small, since only $\sim$10\% of low-mass stars are found in triples \citep[e.g.,][]{2008MNRAS.389..925T,2017ApJS..230...15M,2023ASPC..534..275O}, and the method is effective only for nearby systems, where parallax errors are small enough for reliable matches. For each symbiotic star, we first selected candidates within 20\arcsec{} with parallax measurements with S/N $\geq5$ and consistent with the primary within 3$\sigma$. We then converted proper motions into tangential velocities and compared the velocity vectors, accounting for the uncertainties in both the primary and candidate, which allows for a more physically meaningful assessment of co-motion. Candidate companions were retained if their velocity vectors were consistent with the primary within 3$\sigma$ or had a total tangential velocity difference smaller than 10~km/s. We confirm with high confidence that the symbiotic system CQ Dra has a resolved companion\footnote{We previously reported this in \citet{2022PhDT........34M}.} about 9\arcsec{} away and fainter by $\sim$10.8 mag in $G$, corresponding to a projected separation of $\sim$1600~au. Based on its position in the color–magnitude diagram, the companion is consistent with an M2.5V star. A second case is UV~Aur, an assumed carbon Mira symbiotic\footnote{Its symbiotic status is debated; see \citet{2009AJ....138.1502H} and discussion in \citet{2025A&A...699A.117M}.}, which has a~known late-B companion separated by $\sim$3\arcsec{}. No other convincing cases are found.

\section{Photometric data}\label{sec:photometry}
\begin{figure*}
\centering
\includegraphics[width=\textwidth]{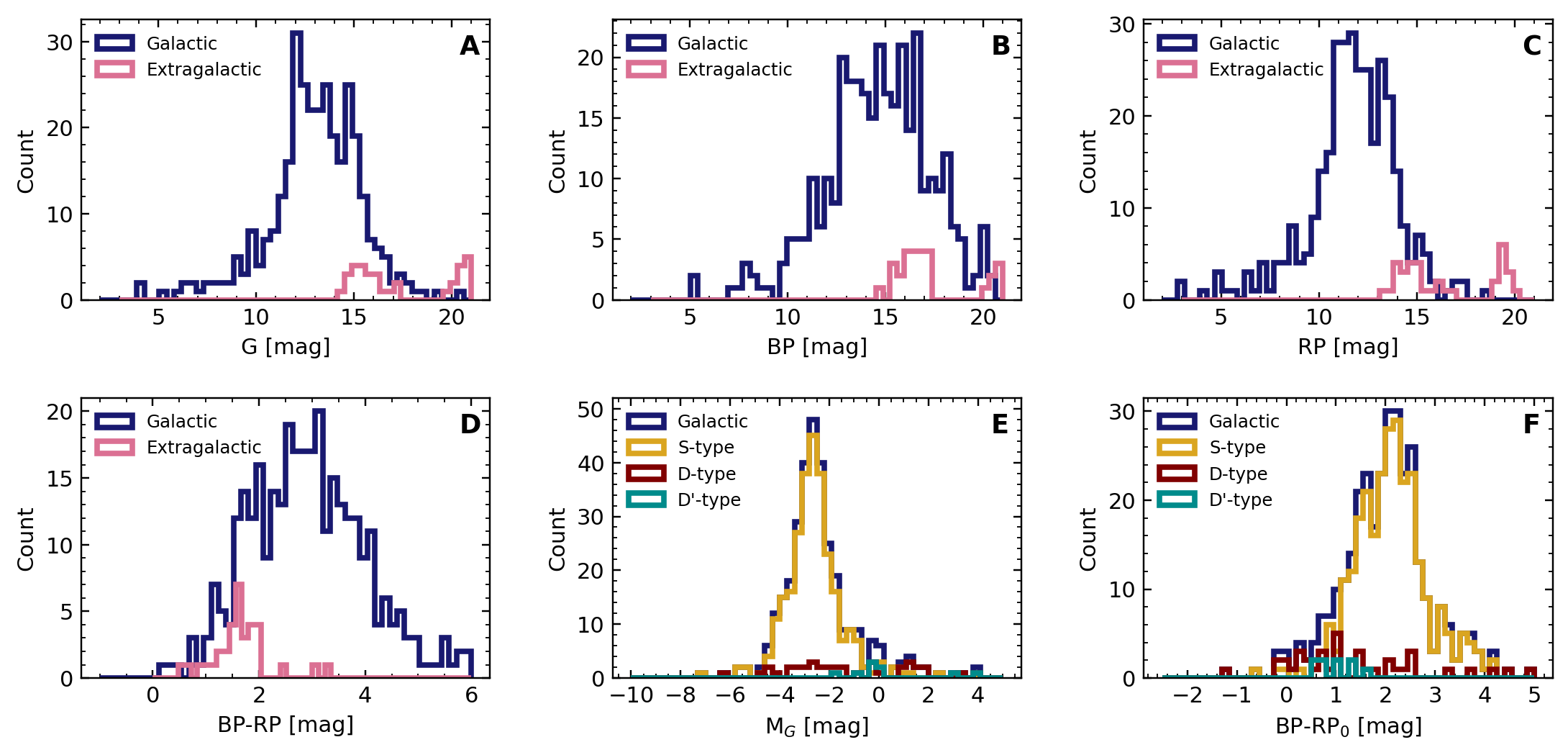}
\caption{Mean photometry of symbiotic stars. \textbf{A:} Distribution of mean apparent $G$ magnitudes of known Galactic and extragalactic symbiotic stars.
\textbf{B:}~Distribution of mean apparent $BP$ magnitudes.
\textbf{C:} Distribution of mean apparent $RP$ magnitudes.
\textbf{D:} Distribution of mean observed $BP-RP$ colors.
\textbf{E:}~Absolute $G$-band magnitudes of symbiotic stars, color-coded by IR type (S, D, D'). Values are calculated using geometric distances from \citet{2021AJ....161..147B} and corrected for extinction using 3D dust maps (see text).
\textbf{F:} Interstellar extinction–corrected $BP-RP$ colors of symbiotic stars.}
\label{fig:mags}
\end{figure*}
\subsection{Mean photometry and color-magnitude diagram}
In addition to astrometric data, \textit{Gaia} provides a wide range of photometric information. The main DR3 catalog lists mean magnitudes derived from varying numbers of observations (CCD transits), depending on the scanning law: between 41 and 1057 in $G$, 2 and 114 in $BP$, and 3 and 115 in $RP$ (although the number of observations used for variability studies was lower, see below). Figures~\ref{fig:mags}A, \ref{fig:mags}B, and \ref{fig:mags}C show the distribution of mean apparent magnitudes for symbiotic stars, separating Galactic systems from extragalactic ones. Besides stars in the Milky Way satellites (LMC, SMC, Draco Dwarf), we also include those in more distant galaxies (e.g., M31 and M33). Symbiotics in satellites cluster around \textit{G}$\sim$15--17 mag, while those in more distant galaxies lie at the faint end of detection limits of \textit{Gaia}. Distribution of observed $BP-RP$ colors is shown in Fig.~\ref{fig:mags}D.

We adopted the geometric distances from \citet{2021AJ....161..147B} and extinction values derived using the {\tt mwdust} code \citep{2016ApJ...818..130B} from combined 3D dust maps of \citet{2003A&A...409..205D}, \citet{2006A&A...453..635M}, and \citet{2019ApJ...887...93G}, to calculate extinction-corrected $BP-RP$ colors and absolute magnitudes $M_G$. Their distributions are shown in Figs.~\ref{fig:mags}E and \ref{fig:mags}F\footnote{In Fig.~\ref{fig:extinc}, we compare the $G$-band extinction from 3D dust maps with the values published in the {\tt astrophysical\_parameters} table, derived by the GSP-Phot Aeneas best-library solution using $BP/RP$ spectra. We also calculated the distribution of $M_G$ using these extinctions. However, the \textit{Gaia}-based extinction values introduce a much larger scatter.}. The median $M_G$ is -2.59 mag with a standard deviation of 1.70 mag, while the median $BP-RP$ is 2.05 mag with a standard deviation of 0.92 mag. Many of the apparently faintest objects are D-type systems, but it should be noted that the correction accounts only for interstellar extinction, whereas these dusty systems also suffer significant intrinsic extinction.

\begin{figure}
\centering
\includegraphics[width=\columnwidth]{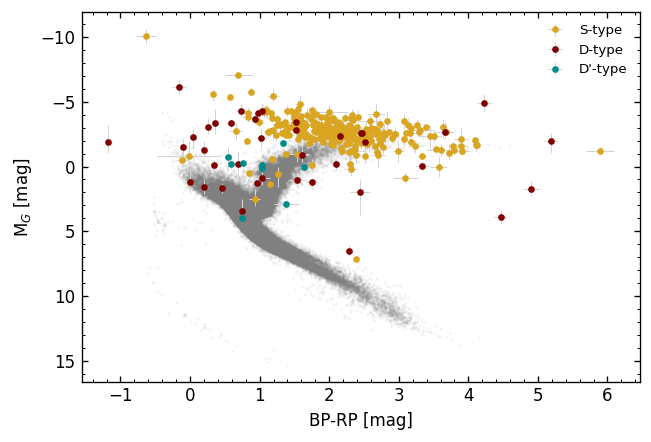}

\caption{Position of symbiotic stars in the \textit{Gaia} color–magnitude diagram. Different IR types (S, D, D') of symbiotics are distinguished by color. The background shows a well-characterized sample of \textit{Kepler} stars from \citet{2025A&A...696A.243G}.   }
\label{fig:hr}
\end{figure}

The corrected values allow us to place symbiotics in the \textit{Gaia} color-magnitude diagram (Fig.~\ref{fig:hr}). For comparison, we also plot the well-characterized stellar sample from \citet{2025A&A...696A.243G}, highlighting major evolutionary sequences such as the main sequence, red giant branch, and white dwarf locus. Most symbiotics lie in the red giant region, as expected, but a non-negligible fraction appear bluer. This partly reflects uncertainties in extinction corrections and distances, but primarily arises because \textit{Gaia} photometry includes a contribution from the nebula (see Fig.~\ref{fig:dr3_filters}), which reduces $BP-RP$ relative to single red giants of the same spectral type. D-type symbiotics also appear bluer. Here the nebular continuum dominates much of the \textit{Gaia} $BP/RP$ passbands, while the cool component is additionally dimmed by self-produced dust. In many symbiotic Miras, dust extinction preferentially obscures the red giant more strongly than the hot companion \citep[e.g.,][]{1999MNRAS.305..190M,2000ASPC..199..431M,Merc+SALT_PaperII}.

Two objects stand out as peculiar in the diagram, with $M_G = 6.5$ and $7.1$ mag. The first is the D-type system JaSt 79, hosting a~Mira pulsator discovered by \citet{2013MNRAS.432.3186M}. Its infrared spectrum \citep{Merc+2025double} clearly confirms its symbiotic Mira nature, and its faint $G$-band absolute magnitude is attributable to additional dust extinction. The second, SWIFT J171951.7-300206, was classified as a~symbiotic by \citet{2012A&A...538A.123M} based on the identification of an optical counterpart to a hard X-ray source, showing an \mbox{M-type} continuum with Balmer and \ion{Ca}{ii} emission. This classification was further supported by \citet{2013A&A...559A...6L}, who analyzed \textit{Swift} observations of the star. However, with a distance of only 187 pc, its luminosity is incompatible with a red giant companion. Based on the same evidence, \citet{2024A&A...690A.243S} recently proposed that the source is instead a very active late-type dwarf, a~conclusion we concur with.

\subsection{Photometric variability}

Out of the 351 symbiotic stars in the Milky Way, LMC, SMC, and Draco Dwarf with \textit{Gaia} counterparts, 331 are classified as variable in \textit{Gaia} DR3\footnote{This does not necessarily imply that the remaining sources are constant; in DR3 the variability flag is either "variable" or "not available," with the latter often due to insufficient data.}. Epoch photometry for these sources is also published in DR3. 

\begin{figure*}
\centering

\includegraphics[width=\textwidth]{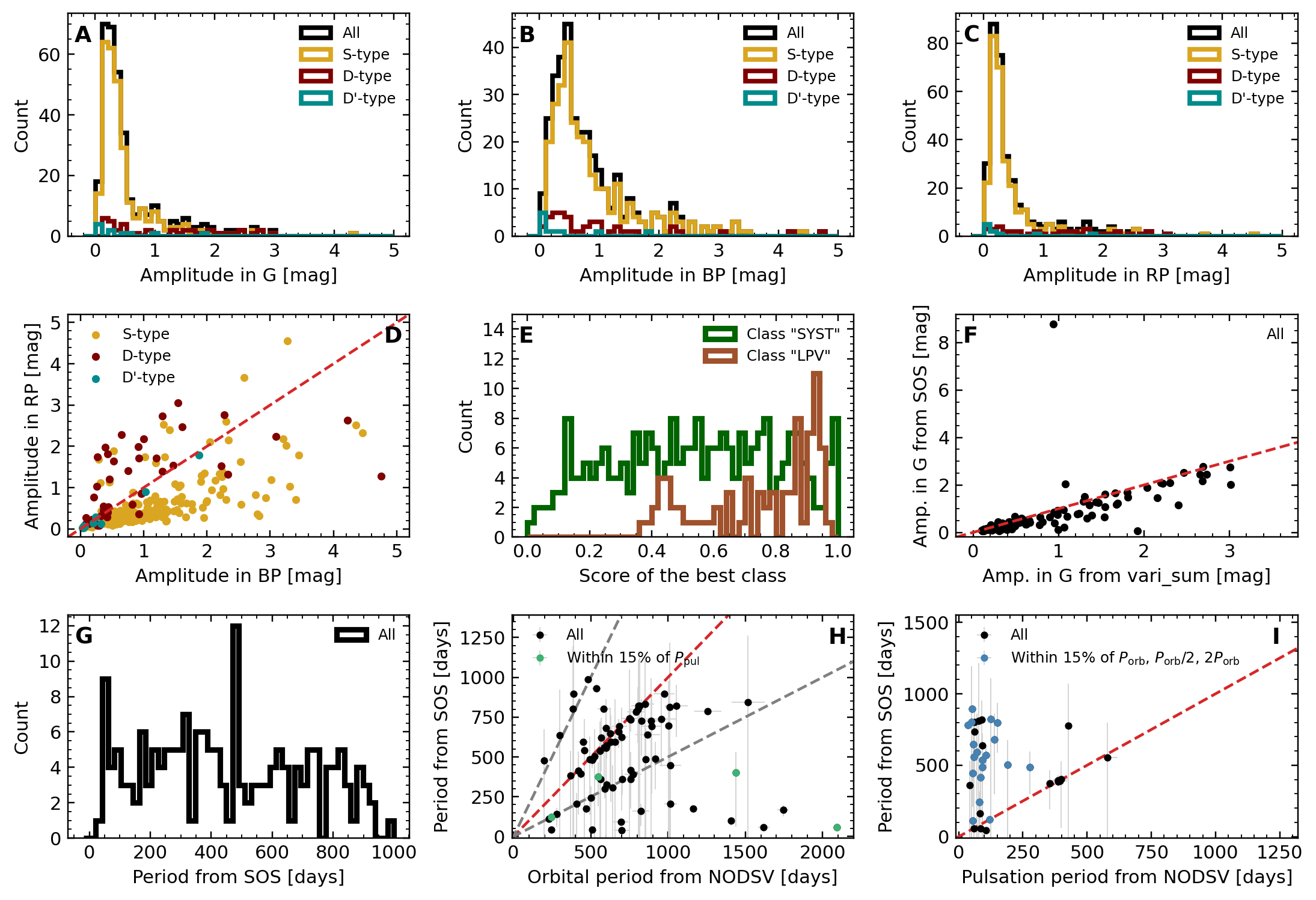}

\caption{Variability of symbiotic stars as seen by \textit{Gaia}.
\textbf{A:} Variability amplitude in the $G$ band from the {\tt vari\_summary} table, color-coded by IR type (S, D, D').
\textbf{B:}~Variability amplitude in the $BP$ band.
\textbf{C:} Variability amplitude in the $RP$ band.
\textbf{D:} Comparison of $BP$ and $RP$ variability amplitudes. The red dashed line indicates the 1:1 relation.
\textbf{E:} Score of the best class from the general variability classification in the {\tt vari\_classifier\_result} table. Only two classes are shown (“symbiotic stars” and “long-period variables”). Five symbiotic stars have a different best class (see text).
\textbf{F:}~Comparison of the $G$-band variability amplitude from the {\tt vari\_summary} table with that derived from the model fit for sources with an inferred period in SOS LPV ({\tt vari\_long\_period\_variable}). The red dashed line indicates the 1:1 relation.
\textbf{G:} Distribution of periods reported in SOS LPV for known symbiotic stars.
\textbf{H:} Comparison of orbital periods from NODSV with periods reported in SOS LPV. The red dashed line indicates the 1:1 relation, while the gray lines indicate 1:2 and 2:1 ratios. Colored points mark cases where the reported SOS period is within 15\% of the pulsation period from NODSV.
\textbf{I:}~Comparison of pulsation periods from NODSV with periods reported in SOS LPV. The red dashed line indicates the 1:1 relation. Colored points mark cases where the reported SOS period is within 15\% of the orbital period, half the orbital period, or twice the orbital period from NODSV.}
\label{fig:variability}
\end{figure*}

Figures~\ref{fig:variability}A, \ref{fig:variability}B, and \ref{fig:variability}C show the distribution of variability amplitudes in $G$, $BP$, and $RP$ filters, as reported in the DR3 {\tt vari\_summary} table\footnote{We note that {\tt vari\_summary} amplitudes that we use are "trimmed," measured between the 5th and 95th percentiles of the magnitude distribution of the $G$-band time series.}. The median amplitudes are 0.34, 0.60, and 0.28 mag, respectively. The figure also separates sources by infrared class. Notably, S-type symbiotics tend to have larger amplitudes in $BP$ than in $RP$ (Fig.~\ref{fig:variability}D), reflecting variability dominated by orbital effects, which are stronger at shorter wavelengths. Conversely, D-type symbiotics, whose variability is dominated by Mira pulsations, show the opposite trend. Some S-type symbiotics in this region have also been reported to host Mira donors, although these systems are not as dusty as D-type symbiotic Miras.

In DR3, statistical and supervised machine-learning methods were used to classify variables into 35 types and subtypes \citep[via the general variability pipeline; see][]{2023A&A...674A..13E,2023A&A...674A..14R}. For symbiotics, the training set \citep[see][]{2023A&A...674A..22G} was mainly drawn from \citet{2000A&AS..146..407B} and \citet{2019ApJS..240...21A}, and classification was based on parameters of \textit{Gaia} light curves and position in the color-magnitude diagram \citep[see the exact list of attributes used in classification is listed in Appendix B of][]{2023A&A...674A..14R}. In total, 649 sources are classified as "symbiotic," (see {\tt vari\_classifier\_result} table) including known symbiotic stars, previously identified candidates (consequently listed in NODSV), and roughly half that are entirely new candidates. Analysis of this full sample is beyond the scope of this work and will be presented in Paper II \citep[][in prep.]{Merc+Gaia2}. 

Among the known symbiotics discussed here, 247 are classified as symbiotic, 79 as long-period variables (LPVs), three as RS~Canum Venaticorum variables (two in the SMC and Hen 3-1591), and two as belonging to a class including B-type emission-line stars, $\gamma$ Cas, S Doradus, or Wolf-Rayet stars (CN~Cha, Hen 3-1613). DR3 only publishes the best-class score from combined classifiers in the {\tt vari\_classifier\_result} table. Figure~\ref{fig:variability}E shows the score distribution for objects classified as symbiotics and LPVs. Several known symbiotics, including those in the training set, are classified with low scores \citep[but were kept, see Sect 3.5 in][]{2023A&A...674A..14R}, highlighting the difficulty of classifying symbiotics based solely on photometry, color, and luminosity \citep[see also][]{2025Galax..13...49M}. This has direct implications for newly discovered candidates.

Some variability classes were further analyzed via the "specific object studies" (SOS), including LPVs \citep[][]{2023A&A...674A..15L}. From the known symbiotics, 278 are included in the SOS LPV list ({\tt vari\_long\_period\_variable}). There is significant overlap between SOS and the general classification. Only 522 of the 1\,720\,588 SOS LPVs were not classified as LPVs in the general pipeline, however, not in all cases, LPV was published as the best class. This explains why only 79 symbiotics appear as LPVs in the general classification, whereas more are included in the SOS LPV table. SOS also provides additional parameters such as dominant frequency, amplitude, and carbon-star candidates, identified through the separations of the two highest peaks in $RP$ spectra, discussed further in Sect.~\ref{sec:carbon}.

Figure~\ref{fig:variability}F compares $G$-band amplitudes from the {\tt vari\_summary} and SOS tables. SOS amplitudes, derived from model fits to objects with periods\footnote{The {\tt vari\_long\_period\_variable} table itself lists half of the peak-to-peak amplitude.}, generally agree with {\tt vari\_summary} values, except in a few cases. A significant outlier is H 2-38, a D-type symbiotic Mira where SOS lists 8.8 mag, while {\tt vari\_summary} reports 0.9 mag. A~review of the $G$-band light curve reveals clear periodic variability between 12.0 and 13.1 mag, indicating that the SOS amplitude is incorrect. The SOS period of 375$\pm$9 days, on the other hand, is close to the literature value of 395$\pm$16 days \citep{2009AcA....59..169G}.

In Fig.~\ref{fig:variability}G, we show the distribution of periods detected by the SOS package for 178 sources, ranging from 40 to 987 days, which encompass the typical pulsational and orbital timescales of symbiotic stars \citep[see Figs. 2 and 3 in][]{Merc+NODSV2025}. Of particular interest is the comparison with known orbital\footnote{Many of the NODSV orbital periods are inferred from photometry only.} and pulsation periods from NODSV. A total of 82 symbiotic stars have an orbital period listed in NODSV and a frequency in SOS (Fig.~\ref{fig:variability}H). Among these, 28 periods agree within 15\% of the NODSV orbital period, 5 are close to twice the orbital period, and 16 are near half of the orbital period. This suggests that a~fraction of symbiotic stars may exhibit ellipsoidal variability detectable by \textit{Gaia}. Additionally, 62 stars have a pulsation period listed in NODSV and also the SOS frequency, of which 24 agree within 15\% of the literature value (Fig.~\ref{fig:variability}I). This comparison demonstrates that \textit{Gaia} light curves of symbiotic stars reflect a combination of orbital variability and pulsations. Consequently, variability classification in \textit{Gaia} DR3 is challenging, as it is often difficult to distinguish symbiotic stars from single pulsating giants using the data employed in the classification.

\section{Spectroscopic data}\label{sec:spectroscopy}

\begin{figure}
\centering
\includegraphics[width=\columnwidth]{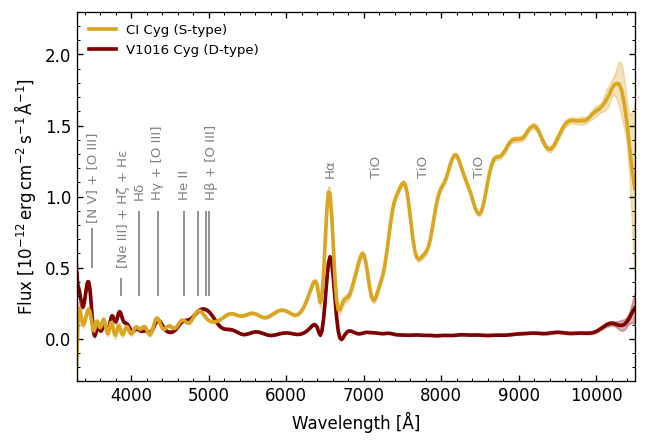}

\caption{Typical $BP/RP$ spectra of S-type (CI Cyg) and D-type (V1016~Cyg) symbiotic stars, with the main emission and absorption features indicated.}
\label{fig:bprp_spec}
\end{figure}

\begin{figure}
\centering
\includegraphics[width=\columnwidth]{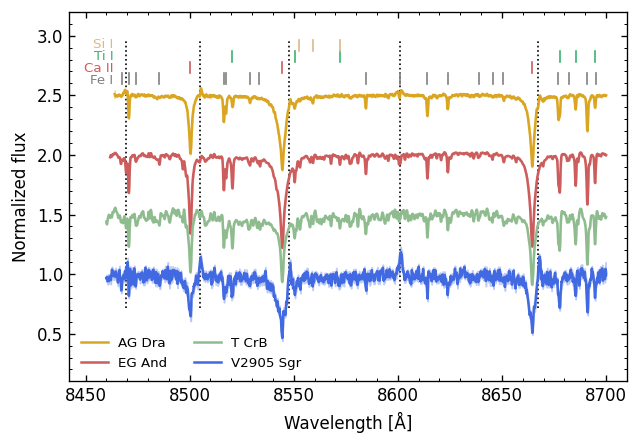}

\caption{RVS spectra of selected symbiotic stars. Absorption features are marked with solid gray lines \citep[][]{2021A&A...654A.130C}, while emission lines of the Paschen \ion{H}{i} series are indicated with dotted black lines.}
\label{fig:rvs_spec}
\end{figure}

In principle, \textit{Gaia} provides two types of spectroscopic data. The low-resolution $BP/RP$ spectra ($R \sim 20$--60) cover the optical to near-infrared region. Typical $BP/RP$ spectra of S- and D-type symbiotic stars are shown in Fig.~\ref{fig:bprp_spec}. The medium-resolution spectra are obtained with the Radial Velocity Spectrometer (RVS; $R \sim 11\,500$) and sample the narrow interval in near-infrared (8\,450 –- 8\,720 \AA{}). The RVS spectra of selected symbiotic stars, with absorption and emission features marked, are shown in Fig.~\ref{fig:rvs_spec}. Beyond the mean spectra themselves, a number of parameters and diagnostics are derived from these observations, some of which we discuss below.

\subsection{Radial velocities and orbits}
\begin{figure*}
\centering
\includegraphics[width=\textwidth]{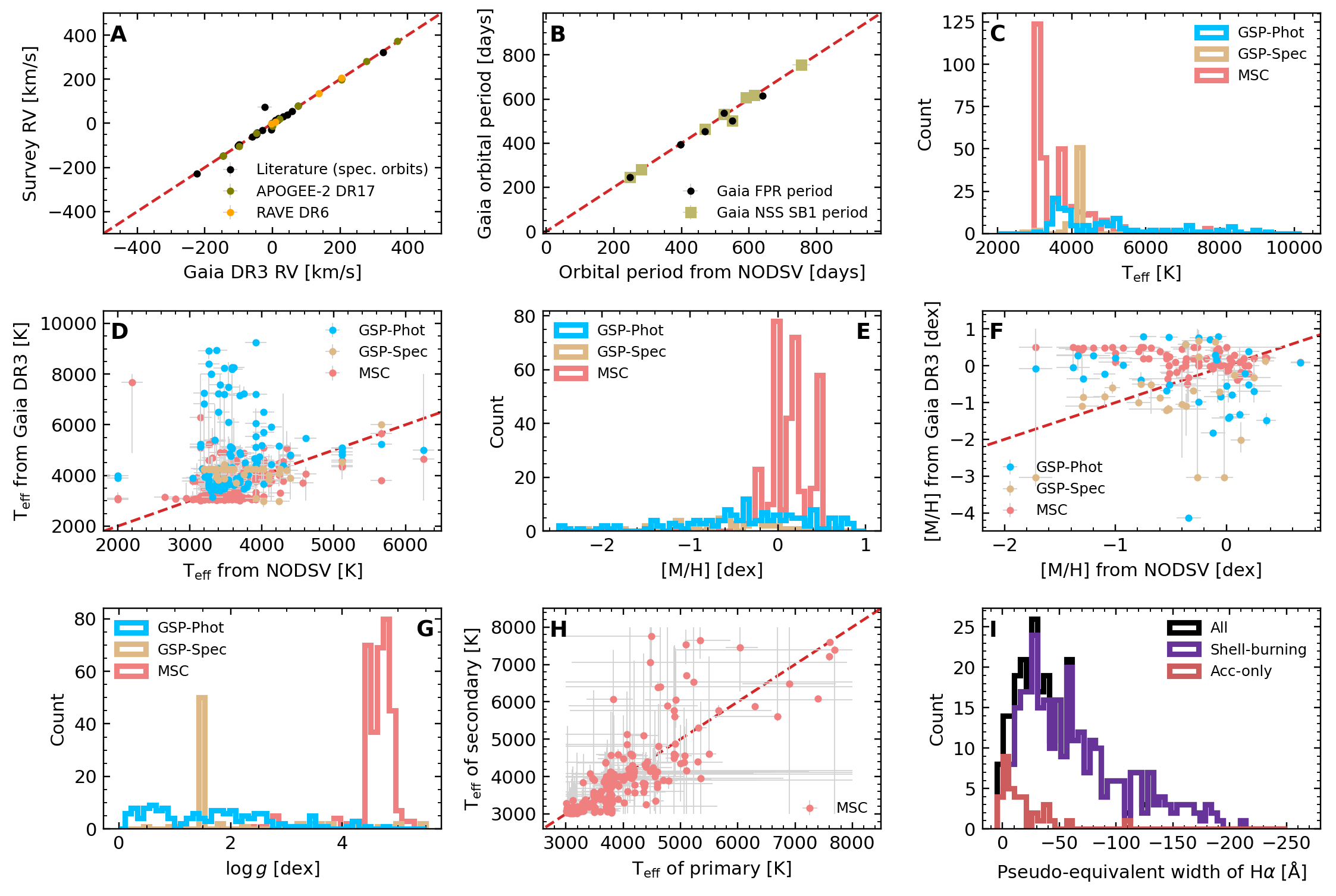}

\caption{Spectroscopic information on symbiotic stars from \textit{Gaia} DR3.
\textbf{A:} Comparison of mean DR3 radial velocities with literature values ($\gamma$ velocities from spectroscopic orbits) and with values published in APOGEE-2 DR17 \citep[][]{2022ApJS..259...35A} and RAVE DR6 \citep[][]{2020AJ....160...82S}.
\textbf{B:} Comparison of orbital periods of symbiotic stars from NODSV with RV variability periods published in FPR \citep[][]{2023A&A...680A..36G} and with orbital periods from NSS SB1 solutions in the {\tt nss\_two\_body\_orbit} table.
\textbf{C:} Distribution of $T_{\rm eff}$ from the GSP-Phot, GSP-Spec, and MSC pipelines. Extreme values from GSP-Phot (up to 37\,000 K) are not shown.
\textbf{D:} Comparison of $T_{\rm eff}$ from NODSV with values from the three \textit{Gaia} pipelines. Extreme GSP-Phot values (up to 37\,000 K) are not shown.
\textbf{E:} Distribution of metallicities from the GSP-Phot, GSP-Spec, and MSC pipelines.
\textbf{F:} Comparison of literature metallicities (from NODSV) with values from the three \textit{Gaia} pipelines.
\textbf{G:} Distribution of $\log g$ from the GSP-Phot, GSP-Spec, and MSC pipelines.
\textbf{H:} Comparison of $T_{\rm eff}$ of the primary and secondary components from the MSC pipeline.
\textbf{I:}~Distribution of pseudo-equivalent widths of H$\alpha$ from the ESP-ELS algorithm, with colors separating shell-burning and accreting-only symbiotic stars.}
\label{fig:spectra}
\end{figure*}

Median radial velocities are available for 130 Galactic symbiotic stars, each based on between 2 and 24 visibility periods \citep[][]{2023A&A...674A...5K}. A comparison of these values with literature $\gamma$-velocities from known spectroscopic orbits, as well as with independent measurements from large surveys such as the Apache Point Observatory Galactic Evolution Experiment 2 \citep[APOGEE-2 DR17;][]{2022ApJS..259...35A} and the Radial Velocity Experiment \citep[RAVE DR6;][]{2020AJ....160...82S}, shows very good agreement \citep[Fig.~\ref{fig:spectra}A; see also][]{2025A&A...698A.155L}. The only clear outlier is T~CrB, which also has the highest RV goodness-of-fit value, suggesting that its \textit{Gaia} solution is unreliable.

As already noted, no epoch radial velocities were published in DR3, but 9 known symbiotic stars were included in the \textit{Gaia} FPR \citep{2023A&A...680A..36G}, each with 14--28 individual RVs spanning 693--969 days. Several of these objects have orbital periods reported in the literature (V1261 Ori\footnote{The FPR RVs of V1261~Ori were further combined with archival data by \citet{2025A&A...695A..61M} to refine the spectroscopic orbit of the system.}, AG~Dra, Hen 3-1213, StHA~176, V417 CMa, LT Del), and the periods listed in the FPR are broadly consistent with those values (Fig.~\ref{fig:spectra}B). Others (StHA 154, StHA 151) have no spectroscopic or photometric periods reported. Using the {\tt radial} code\footnote{\url{https://github.com/ladsantos/radial}}, we analyzed their FPR RVs and obtained the first preliminary orbital constraints for these objects, presented in Table~\ref{tab:orbit} and illustrated in Fig.~\ref{fig:orbits}. As emphasized by \citet{2025A&A...695A..61M} in the case of V1261 Ori, the DR3 timespan is still too short to yield fully robust orbital solutions, but significant improvements are expected with \textit{Gaia} DR4.

Interestingly, the RV periods reported in the FPR for V1261 Ori, V417 CMa, RT Ser, and StHA 176 are very close to twice the photometric periods listed in the FPR/SOS LPV, consistent with the presence of ellipsoidal variability in these symbiotic systems. The case of the symbiotic nova RT~Ser is particularly intriguing. A period of about 12 years has been suggested from long-term light curve and H$\alpha$ profile variations \citep[][]{2003ASPC..303...87S}, whereas the FPR data indicate an RV period of 355 days. Notably, its photometric period is again close to half the spectroscopic one. Both the RV curve and light curve appear complex, and while a full analysis is beyond the scope of this work, the system clearly merits detailed follow-up.

Apart from the RVs themselves, \textit{Gaia} DR3 also included the non-single star (NSS) catalog, which provides binary parameters inferred from spectroscopy or astrometry. Ten symbiotic stars appear as single-lined spectroscopic binaries (SB1) in the {\tt nss\_two\_body\_orbit} table, while none are listed in other NSS solution types (SB2, acceleration solutions, etc.). As we discussed in \citet{2024A&A...682A...7B}, most of these orbital solutions are consistent with spectroscopic orbits reported in the literature, with only minor deviations for periods longer than $\sim$500 days (Fig.~\ref{fig:spectra}B), likely due to the limited DR3 time coverage of at most two orbital cycles.

One clear spurious case is Hen 3-1308, for which the {\tt nss\_two\_body\_orbit} table lists a period of 0.43 days, an entirely unphysical value for a system containing an evolved red giant donor, and thus a failed solution. No orbital period is available in the literature for this star, nor for II Vul. Four of the SB1 solutions are in common with sources also included in the \textit{Gaia} FPR.

\subsection{Astrophysical parameters}
Both the $BP/RP$ and RVS spectra were used in \textit{Gaia} DR3 to infer astrophysical parameters of sources \citep{2023A&A...674A..26C,2023A&A...674A..28F, 2023A&A...674A..27A,2023A&A...674A..29R}, with the results provided mainly in the tables {\tt astrophysical\_parameters} and {\tt astrophysical\_parameters\_supp}. In this section, we compare effective temperatures ($T_{\rm eff}$) and metallicities derived from \textit{Gaia} DR3 pipelines with values listed in NODSV, which are based primarily on high-resolution infrared spectroscopy, with some of the literature temperature estimates in NODSV relying on spectral-type calibrations only.

Figure~\ref{fig:spectra}C illustrates the comparison of effective temperatures obtained from several approaches: {\tt teff\_gspphot}, from the \textit{General Stellar Parametrizer from Photometry} \citep[\mbox{GSP-Phot}, Aeneas best library;][]{2023A&A...674A..27A}, which uses $BP/RP$ spectra, $G$ magnitudes, and parallaxes under the assumption of single stars; {\tt teff\_gspspec}, from the \textit{General Stellar Parametrizer from Spectroscopy} (\mbox{GSP-Spec}; \citealt{2023A&A...674A..29R}), which derives parameters from RVS spectra and Monte Carlo realizations; {\tt teff\_msc1}, from the \textit{Multiple Star Classifier} (MSC), which infers stellar parameters from $BP/RP$ spectra under the assumption of an unresolved coeval binary. 

The results show clear systematics. GSP-Spec and MSC return reasonable ranges of effective temperatures, whereas GSP-Phot often fails, with some symbiotic stars assigned unrealistic values as high as 37\,000 K (Fig.~\ref{fig:spectra}C). On an object-by-object basis, the derived temperatures are generally unreliable (Fig.~\ref{fig:spectra}D). For metallicities (Fig.~\ref{fig:spectra}E,F), individual values show large scatter and the overall distributions differ from those expected for the known population: the median [Fe/H] of M-type giants in symbiotic stars is $-0.18$ dex, while yellow symbiotic stars show [Fe/H] $\approx -1.15$ dex \citep[][]{Merc+NODSV2025}. Similarly, the distributions of $\log g$ (Fig.~\ref{fig:spectra}G) are inconsistent with evolved red giants. In particular, the MSC pipeline often predicts $\log g$ between 4 and 5, values typical of dwarf stars rather than giants. 

Another noteworthy observation is the comparison between the MSC effective temperatures of the primary and secondary stars: in the catalog, these appear similar (Fig.~\ref{fig:spectra}H), whereas in reality, the components of symbiotic systems differ drastically. Since the hot companions are too hot to contribute significantly to the $BP/RP$ spectra, this outcome is understandable and highlights that the MSC pipeline is not suited to symbiotic binaries.

All these complications likely arise from the symbiotic nature of the targets: being emission-line sources with strong nebular continua, their observed spectra are not those of the cool giants alone (for example \ion{Ca}{ii} lines in the RVS spectrum are contaminated by Paschen \ion{H}{i} emission lines, while continuum in $BP/RP$ spectrum is affected by strong H$\alpha$ and other emission lines; see Figs. \ref{fig:bprp_spec} and \ref{fig:rvs_spec}). 

\subsection{H$\alpha$ emission and emission-line stars classification}
The $BP/RP$ spectra were analyzed by the \textit{Extended Stellar Parametrizer for Emission-Line Stars} (ESP-ELS) algorithm \citep{2023A&A...674A..26C,2023A&A...674A..28F} to identify and classify emission-line stars into seven categories (Be stars, Herbig Ae/Be stars, T~Tauri stars, active M dwarfs, Wolf–Rayet WC and WN stars, and planetary nebulae). The pipeline processes targets brighter than $G = 17.65$~mag, with the selection based primarily on the pseudo-equivalent width (pEW) of the H$\alpha$ line measured in the $BP/RP$ spectra, complemented by machine-learning methods.

In total, \textit{Gaia} DR3 provides H$\alpha$ emission measurements for 235 million sources and emission-line classifications for 57\,511 stars (table {\tt astrophysical\_parameters}). Among the known symbiotic stars, 342 are brighter than $G = 17.65$ mag, and all of these have published H$\alpha$ pEW values. As shown in Fig.~\ref{fig:spectra}I, virtually all symbiotic stars are detected as strong H$\alpha$ emitters\footnote{Surveys searching for symbiotic stars preferentially select strong emitters by design. Consequently, systems with weak or absent optical emission lines are underrepresented in current catalogs; see, e.g., \citet{2016MNRAS.461L...1M} and the discussion in \citet{2025Galax..13...49M}.}, with shell-burning systems showing systematically stronger emission than accretion-powered systems, as expected.

The main outlier in the distribution of pEW among accreting-only symbiotic stars (which typically show weak or no emission lines) is the neutron-star symbiotic binary V2116~Oph. This system is known to exhibit strong emission lines, including faint Raman-scattered \ion{O}{vi} features, due to the UV radiation from the accretion disk \citep{1997ApJ...489..254C}.

While a direct comparison of the H$\alpha$ strength measured by \textit{Gaia} with literature values is complicated by the strong variability of symbiotic stars, we compared the \textit{Gaia} pEW(H$\alpha$) with average EW values from spectra collected in the Astronomical Ring for Amateur Spectroscopy (ARAS) database \citep[][]{2019CoSka..49..217T} to obtain at least a qualitative assessment. The comparison, shown in Fig.~\ref{fig:halpha_ARAS}, reveals a reasonable correlation despite the fact that we did not restrict the ARAS spectra to the same time interval as the \textit{Gaia} DR3 observations, and despite the strong dependence of H$\alpha$ strength on orbital phase and activity state. We performed a linear fit constrained to pass through (0,0), limited to ARAS EWs larger than $-400$~\AA{} where the correlation is strongest. This first-order estimate indicates that the \textit{Gaia} pEW values are typically about 0.3 times the ARAS measurements. Very strong emission lines, especially in some D-type symbiotic systems, are significantly underestimated by \textit{Gaia}.

In any case, the analysis of pEW from \textit{Gaia} DR3 clearly demonstrates that the \textit{Gaia} $BP/RP$ spectra are fully adequate to reveal the emission-line nature of symbiotic stars, and thus represent a powerful tool for the analysis of candidates and the search for new members of this class.

Among the symbiotic stars analyzed with the ESP-ELS pipeline, 195 received an emission-line star classification. It is important to note that no dedicated symbiotic star class was available; therefore, all of these objects are formally misclassified. Of these, 168 were classified as T Tauri stars, with probabilities between 0.20 and 0.95. This is not unexpected, as T Tauri stars are common contaminants in symbiotic samples: their continua are dominated by cool stars, and they exhibit strong emission lines. A further 24 were classified as planetary nebulae, with probabilities between 0.43 and 0.91. With one exception, all of these correspond to D- or D'-type symbiotic stars in the literature, where dust extinction obscures the cool component in the $BP/RP$ spectra, leaving a continuum that resembles that of planetary nebulae. Such misidentifications are not surprising, as many known symbiotic stars were historically catalogued as PNe and were only reclassified once infrared data became available. Three objects (RX Pup, JaSt 79, and K~3-22, all D-types with virtually no continuum in $BP/RP$ spectra) were classified as Herbig Ae/Be stars, with probabilities between 0.25 and 0.67.

This exercise highlights that additional symbiotic stars may be hidden among the emission-line star classifications in \textit{Gaia} DR3. In particular, new D-type symbiotic stars (which are generally difficult to identify) may be found among planetary nebula candidates \citep[][]{Mulato+PNe}.

\subsection{Carbon classification from SOS LPVs}\label{sec:carbon}
In the SOS LPV package \citep[][]{2023A&A...674A..15L}, the $RP$ spectra were used to identify potential carbon star candidates. The {\tt vari\_long\_period\_variable} table includes, in addition to the carbon star candidate flag ({\tt isCstar} = 0, 1, or empty), the median value of the pseudo-wavelength difference between the two highest peaks in the $RP$ spectrum of each source ({\tt median\_delta\_wl\_rp}). While all stars with a difference greater than 7 are flagged as carbon-rich, the authors recommend selecting candidates with \mbox{$7 < {\tt median\_delta\_wl\_rp} < 11$} and with $BP < 19$ mag for more reliable identification.

Selecting stars with the {\tt isCstar} flag equal to 1 yields 88 objects, while 188 are classified as O-rich. This fraction is much larger than the number of known carbon symbiotic stars \citep[11 in the Milky Way, one in the Draco Dwarf, NGC 6822, four in the LMC, two in the SMC, and four in M33; e.g.,][]{2017MNRAS.465.1699M,2020ApJ...900L..43L,2025A&A...699A.117M}. Applying the recommended range for the pseudo-wavelength separation ($7 < {\tt median\_delta\_wl\_rp} < 11$) reduces the sample to 14 objects. Of these, nine are genuine carbon symbiotic stars (including three in the LMC), while five have \mbox{an O-rich} classification in the literature: EG And, IGR J17329-2731, [MMU2013] 357.98+01.57, 2MASS J17374702-2501120, and \mbox{PN~K~5-8}. We reviewed their $BP/RP$ spectra (externally calibrated mean data) and found that, except for \mbox{PN~K~5-8}, whose unusual spectrum is likely affected by observational artifacts, the others show maxima expected for M-type stars. We also employed internally calibrated spectra of these five stars, which allowed us to measure the same quantity, i.e., the pseudo-wavelength separation between the two highest peaks in the RP spectrum, as reported in DR3. In all five cases, our measured values are consistent with an O-rich rather than a C-rich classification. The key difference between our {\tt delta\_wl\_rp} and the DR3 value is that ours is derived from the mean RP spectrum, whereas the DR3 value was computed from epoch spectra (which are not publicly available) and the median of those measurements was published. In these particular cases, the mean-spectrum measurement appears to provide a more robust result than the epoch-based median.

Only V1196 Sco and SS73 38 are misclassified as O-rich in DR3, but the {\tt median\_delta\_wl\_rp} values for both are negative. \mbox{H 1-45} is flagged as a carbon candidate, but its separation ($\sim$24) lies well outside the recommended range. All three objects are D-type symbiotic stars, with spectra dominated by very strong emission lines that overshine the cool giant. Finally, the remaining known carbon symbiotic stars are not included in the SOS LPV table. 

\section{Conclusions}\label{sec:conclusions}
In this work, we examined the information available for known symbiotic stars in \textit{Gaia} DR3 and FPR. We analyzed astrometric, photometric, spectroscopic, and variability data, together with additional inferred parameters (orbits, emission-line classifications, etc.). Our main conclusions are as follows:\\

\noindent \textbf{Astrometry, distances, and photometry}
\begin{itemize}
    \item Many symbiotic stars have poorly measured parallaxes; nevertheless, their astrometric data are sufficient to place them in the color–magnitude diagram and separate them from, e.g., cool main-sequence stars, although they are maybe not yet precise enough to accurately determine distance-dependent parameters such as luminosity or the radius of the giant.

 \item Using \textit{Gaia}-based distances, we computed vertical distances from the Galactic plane. The distribution of S-type symbiotic stars shows an intriguing bimodality, with a slight deficit near $z \sim 0$.

 \item Two systems (CQ Dra and UV Aur) show resolved companions in \textit{Gaia}, suggesting they may be triples.

 \item In some objects, additional nebular emission shifts their position blueward of the expected red-giant locus in the color–magnitude diagram.

 \item Based on its astrometric and photometric properties, we reclassify SWIFT J171951.7$-$300206 as a non-symbiotic, active M-type dwarf.

 \item RUWE is not a reliable binary indicator for symbiotic stars, making its use in searches for new systems questionable.
\end{itemize}
\vspace{3mm}
\noindent \textbf{Variability and light curves}
\begin{itemize}
    \item The majority of known symbiotic stars are detected as variables by \textit{Gaia}, with a large fraction analyzed within the SOS module for long-period variables.

\item S-type systems typically show larger amplitudes in $BP$ than in $RP$, consistent with variability dominated by orbital motion. In D-type systems, Mira pulsations dominate the light curves and, contrary to single or less dusty Miras \citep[e.g.,][]{2021A&A...648A..44M}, show larger amplitudes in $RP$, due to the combined effects of larger dust extinction and nebular emission in $BP$.

\item Comparison with literature periods shows that both orbital and pulsation periods can dominate the light curves, a fact important for variability classification.
\end{itemize}
\vspace{3mm}
\noindent \textbf{Spectroscopy and orbital solutions}

\begin{itemize}
    \item Mean \textit{Gaia} radial velocities are generally consistent with the literature and are useful for computing 3D velocities.

\item Periods inferred from epoch RVs and NSS SB1 solutions are in reasonable agreement with literature values, with significant improvement expected as the time baseline grows.

\item We provide the first preliminary orbital solutions for two systems using \textit{Gaia} FPR epoch RVs: StHA 154 ($P_{\rm orb} = 571$ d, $e=0.58$) and StHA 151 ($P_{\rm orb} = 668$ d, $e=0.04$).

\item \textit{Gaia} photometric and spectroscopic data indicate the presence of ellipsoidal variability in a fraction of symbiotic stars, warranting further investigation, particularly with a longer timebase.

\item Nearly all symbiotic stars show strong emission lines detectable in $BP/RP$ spectra, especially H$\alpha$, making this a powerful identifier of similar systems.

\item Emission-line classification in \textit{Gaia} did not include a~symbiotic/star class: S-types are mostly misclassified as T~Tauri stars, and D-types as planetary nebulae. Thus, searches among the \textit{Gaia} PNe may reveal additional D-type candidates.

\item Astrophysical parameters ($T_{\rm eff}$, [Fe/H], $\log g$) derived from \textit{Gaia} pipelines are unreliable for symbiotic stars, since both $BP/RP$ and RVS spectra are contaminated by nebular continuum and strong emission lines.

\item Carbon-star classification in the SOS LPV pipeline is partly reliable for symbiotics if specific peak-separation criteria are applied; using only the {\tt isCstar} flag leads to strong contamination.

\end{itemize}

Overall, the \textit{Gaia} DR3 and FPR datasets already provide a~rich resource for the study of symbiotic stars. They enable analysis of known systems, re-assessment of literature candidates, and systematic searches for new members of this class. At the same time, our results highlight both the opportunities and the limitations of the current release. Looking ahead, the prospects for \textit{Gaia} DR4 (scheduled for December 2026) are particularly exciting: the longer time baseline, the expanded set of data products, and the availability of epoch astrometric, photometric, and spectroscopic data for all sources will open entirely new possibilities for characterizing these complex interacting binaries.


\section*{Acknowledgements}

 We thank the anonymous referee for the careful review and helpful suggestions that improved the manuscript. The research of J.M. was supported by the Czech Science Foundation (GACR) project no. 24-10608O and by the Spanish Ministry of Science and Innovation with the grant no. PID2023-146453NB-100 (PLAtoSOnG). This work has made use of data from the European Space Agency (ESA) mission
{\it Gaia} (\url{https://www.cosmos.esa.int/gaia}), processed by the {\it Gaia}
Data Processing and Analysis Consortium (DPAC,
\url{https://www.cosmos.esa.int/web/gaia/dpac/consortium}). Funding for the DPAC
has been provided by national institutions, in particular the institutions
participating in the {\it Gaia} Multilateral Agreement.

\section*{Data Availability}
Data used in this work are available from \textit{Gaia} DR3 archive and the New Online Database of Symbiotic Variables.



\bibliographystyle{mnras}
\bibliography{lit} 

@ARTICLE{2019AN....340..598M,
       author = {{Merc}, J. and {G{\'a}lis}, R. and {Wolf}, M.},
        title = "{New online database of symbiotic variables: Symbiotics in X‑rays}",
      journal = {Astronomische Nachrichten},
         year = 2019,
        month = aug,
       volume = {340},
       number = {7},
        pages = {598-606},
          doi = {10.1002/asna.201913662},
       adsurl = {https://ui.adsabs.harvard.edu/abs/2019AN....340..598M}
}

@ARTICLE{2019RNAAS...3...28M,
       author = {{Merc}, Jaroslav and {G{\'a}lis}, Rudolf and {Wolf}, Marek},
        title = "{First Release of the New Online Database of Symbiotic Variables}",
      journal = {Research Notes of the American Astronomical Society},
         year = 2019,
        month = feb,
       volume = {3},
       number = {2},
          eid = {28},
        pages = {28},
          doi = {10.3847/2515-5172/ab0429},
       adsurl = {https://ui.adsabs.harvard.edu/abs/2019RNAAS...3...28M}
}

@ARTICLE{Merc+NODSV2025,
       author = {{Merc}, Jaroslav and {G{\'a}lis}, Rudolf and {Wolf}, Marek},
        title = "{New Online Database of Symbiotic Variables: Catalog and Statistical Overview of Symbiotic Binaries}",
      journal = {submitted to ApJS},
         year = 2026
}

@ARTICLE{2019ApJ...887...93G,
       author = {{Green}, Gregory M. and {Schlafly}, Edward and {Zucker}, Catherine and {Speagle}, Joshua S. and {Finkbeiner}, Douglas},
        title = "{A 3D Dust Map Based on Gaia, Pan-STARRS 1, and 2MASS}",
      journal = {\apj},
         year = 2019,
        month = dec,
       volume = {887},
       number = {1},
          eid = {93},
        pages = {93},
          doi = {10.3847/1538-4357/ab5362},
archivePrefix = {arXiv},
       eprint = {1905.02734},
 primaryClass = {astro-ph.GA},
       adsurl = {https://ui.adsabs.harvard.edu/abs/2019ApJ...887...93G}
}

@ARTICLE{2021A&A...648A..44M,
       author = {{Mowlavi}, N. and {Rimoldini}, L. and {Evans}, D.~W. and {Riello}, M. and {De Angeli}, F. and {Palaversa}, L. and {Audard}, M. and {Eyer}, L. and {Garcia-Lario}, P. and {Gavras}, P. and {Holl}, B. and {Jevardat de Fombelle}, G. and {Lec{\oe}ur-Ta{\"\i}bi}, I. and {Nienartowicz}, K.},
        title = "{Large-amplitude variables in Gaia Data Release 2. Multi-band variability characterization}",
      journal = {\aap},
         year = 2021,
        month = apr,
       volume = {648},
          eid = {A44},
        pages = {A44},
          doi = {10.1051/0004-6361/202039450},
archivePrefix = {arXiv},
       eprint = {2009.07746},
 primaryClass = {astro-ph.SR},
       adsurl = {https://ui.adsabs.harvard.edu/abs/2021A&A...648A..44M}
}

@ARTICLE{2016ApJ...818..130B,
       author = {{Bovy}, Jo and {Rix}, Hans-Walter and {Green}, Gregory M. and {Schlafly}, Edward F. and {Finkbeiner}, Douglas P.},
        title = "{On Galactic Density Modeling in the Presence of Dust Extinction}",
      journal = {\apj},
         year = 2016,
        month = feb,
       volume = {818},
       number = {2},
          eid = {130},
        pages = {130},
          doi = {10.3847/0004-637X/818/2/130},
archivePrefix = {arXiv},
       eprint = {1509.06751},
 primaryClass = {astro-ph.GA},
       adsurl = {https://ui.adsabs.harvard.edu/abs/2016ApJ...818..130B}
}

@ARTICLE{2006A&A...453..635M,
       author = {{Marshall}, D.~J. and {Robin}, A.~C. and {Reyl{\'e}}, C. and {Schultheis}, M. and {Picaud}, S.},
        title = "{Modelling the Galactic interstellar extinction distribution in three dimensions}",
      journal = {\aap},
         year = 2006,
        month = jul,
       volume = {453},
       number = {2},
        pages = {635-651},
          doi = {10.1051/0004-6361:20053842},
archivePrefix = {arXiv},
       eprint = {astro-ph/0604427},
 primaryClass = {astro-ph},
       adsurl = {https://ui.adsabs.harvard.edu/abs/2006A&A...453..635M}
}

@ARTICLE{2003A&A...409..205D,
       author = {{Drimmel}, R. and {Cabrera-Lavers}, A. and {L{\'o}pez-Corredoira}, M.},
        title = "{A three-dimensional Galactic extinction model}",
      journal = {\aap},
         year = 2003,
        month = oct,
       volume = {409},
        pages = {205-215},
          doi = {10.1051/0004-6361:20031070},
archivePrefix = {arXiv},
       eprint = {astro-ph/0307273},
 primaryClass = {astro-ph},
       adsurl = {https://ui.adsabs.harvard.edu/abs/2003A&A...409..205D}
}

@ARTICLE{2025A&A...696A.243G,
       author = {{Godoy-Rivera}, D. and {Mathur}, S. and {Garc{\'\i}a}, R.~A. and {Pinsonneault}, M.~H. and {Santos}, {\^A}. R.~G. and {Beck}, P.~G. and {Grossmann}, D.~H. and {Schimak}, L. and {Bedell}, M. and {Merc}, J. and {Escorza}, A.},
        title = "{Kepler meets Gaia DR3: Homogeneous extinction-corrected color-magnitude diagram and binary classification}",
      journal = {\aap},
         year = 2025,
        month = apr,
       volume = {696},
          eid = {A243},
        pages = {A243},
          doi = {10.1051/0004-6361/202348735},
archivePrefix = {arXiv},
       eprint = {2501.18719},
 primaryClass = {astro-ph.SR},
       adsurl = {https://ui.adsabs.harvard.edu/abs/2025A&A...696A.243G}
}

@INPROCEEDINGS{2000ASPC..199..431M,
       author = {{Miko{\l}ajewska}, Joanna},
        title = "{Observed Properties of Mass Loss in Symbiotic Binaries}",
    booktitle = {Asymmetrical Planetary Nebulae II: From Origins to Microstructures},
         year = 2000,
       editor = {{Kastner}, J.~H. and {Soker}, N. and {Rappaport}, S.},
       series = {Astronomical Society of the Pacific Conference Series},
       volume = {199},
        month = jan,
        pages = {431},
          doi = {10.48550/arXiv.astro-ph/0001012},
archivePrefix = {arXiv},
       eprint = {astro-ph/0001012},
 primaryClass = {astro-ph},
       adsurl = {https://ui.adsabs.harvard.edu/abs/2000ASPC..199..431M}
}

@ARTICLE{2013MNRAS.432.3186M,
       author = {{Miszalski}, Brent and {Miko{\l}ajewska}, Joanna and {Udalski}, Andrzej},
        title = "{Symbiotic stars and other H{\ensuremath{\alpha}} emission-line stars towards the Galactic bulge}",
      journal = {\mnras},
         year = 2013,
        month = jul,
       volume = {432},
       number = {4},
        pages = {3186-3217},
          doi = {10.1093/mnras/stt673},
archivePrefix = {arXiv},
       eprint = {1305.4863},
 primaryClass = {astro-ph.SR},
       adsurl = {https://ui.adsabs.harvard.edu/abs/2013MNRAS.432.3186M}
}

@article{Merc+2025double,
    author = {{Merc}, J. and {Miko\l{}ajewska}, J. and {Ga\l{}an}, C. and {I\l{}kiewicz}, K. and {Beck}, P. G. and {Monard}, B. and {Gromadzki}, M.},
    title = "{Blending-induced beating and emission in the symbiotic star Terz V 2513}",
    journal = {submitted to MNRAS},
    year = 2025
}

@ARTICLE{2012A&A...538A.123M,
       author = {{Masetti}, N. and {Parisi}, P. and {Jim{\'e}nez-Bail{\'o}n}, E. and {Palazzi}, E. and {Chavushyan}, V. and {Bassani}, L. and {Bazzano}, A. and {Bird}, A.~J. and {Dean}, A.~J. and {Galaz}, G. and {Landi}, R. and {Malizia}, A. and {Minniti}, D. and {Morelli}, L. and {Schiavone}, F. and {Stephen}, J.~B. and {Ubertini}, P.},
        title = "{Unveiling the nature of INTEGRAL objects through optical spectroscopy. IX. Twenty two more identifications, and a glance into the far hard X-ray Universe}",
      journal = {\aap},
         year = 2012,
        month = feb,
       volume = {538},
          eid = {A123},
        pages = {A123},
          doi = {10.1051/0004-6361/201118559},
archivePrefix = {arXiv},
       eprint = {1201.1906},
 primaryClass = {astro-ph.HE},
       adsurl = {https://ui.adsabs.harvard.edu/abs/2012A&A...538A.123M}
}

@ARTICLE{2023A&A...674A..15L,
       author = {{Lebzelter}, T. and {Mowlavi}, N. and {Lecoeur-Taibi}, I. and {Trabucchi}, M. and {Audard}, M. and {Garc{\'\i}a-Lario}, P. and {Gavras}, P. and {Holl}, B. and {Jevardat de Fombelle}, G. and {Nienartowicz}, K. and {Rimoldini}, L. and {Eyer}, L.},
        title = "{Gaia Data Release 3. The second Gaia catalogue of long-period variable candidates}",
      journal = {\aap},
         year = 2023,
        month = jun,
       volume = {674},
          eid = {A15},
        pages = {A15},
          doi = {10.1051/0004-6361/202244241},
archivePrefix = {arXiv},
       eprint = {2206.05745},
 primaryClass = {astro-ph.SR},
       adsurl = {https://ui.adsabs.harvard.edu/abs/2023A&A...674A..15L}
}

@ARTICLE{2009AcA....59..169G,
       author = {{Gromadzki}, M. and {Miko{\l}ajewska}, J. and {Whitelock}, P. and {Marang}, F.},
        title = "{Light Curves of Symbiotic Stars in Massive Photometric Surveys I: D-Type Systems}",
      journal = {\actaa},
         year = 2009,
        month = jun,
       volume = {59},
       number = {2},
        pages = {169-191},
          doi = {10.48550/arXiv.0906.4136},
archivePrefix = {arXiv},
       eprint = {0906.4136},
 primaryClass = {astro-ph.SR},
       adsurl = {https://ui.adsabs.harvard.edu/abs/2009AcA....59..169G}
}

@ARTICLE{2020ApJ...900L..43L,
       author = {{Lewis}, Hannah M. and {Anguiano}, Borja and {Stassun}, Keivan G. and {Majewski}, Steven R. and {Arras}, Phil and {Sarazin}, Craig L. and {Li}, Zhi-Yun and {De Lee}, Nathan and {Troup}, Nicholas W. and {Allende Prieto}, Carlos and {Badenes}, Carles and {Cunha}, Katia and {Garc{\'\i}a-Hern{\'a}ndez}, D.~A. and {Nidever}, David L. and {Palicio}, Pedro A. and {Simon}, Joshua D. and {Smith}, Verne V.},
        title = "{Geometry of the Draco C1 Symbiotic Binary}",
      journal = {\apjl},
         year = 2020,
        month = sep,
       volume = {900},
       number = {2},
          eid = {L43},
        pages = {L43},
          doi = {10.3847/2041-8213/abb248},
archivePrefix = {arXiv},
       eprint = {2008.05962},
 primaryClass = {astro-ph.SR},
       adsurl = {https://ui.adsabs.harvard.edu/abs/2020ApJ...900L..43L}
}

@ARTICLE{2017MNRAS.465.1699M,
       author = {{Miko{\l}ajewska}, Joanna and {Shara}, Michael M. and {Caldwell}, Nelson and {I{\l}kiewicz}, Krystian and {Zurek}, David},
        title = "{A survey of the Local Group of galaxies for symbiotic binary stars - I. First detection of symbiotic stars in M33}",
      journal = {\mnras},
         year = 2017,
        month = feb,
       volume = {465},
       number = {2},
        pages = {1699-1710},
          doi = {10.1093/mnras/stw2937},
archivePrefix = {arXiv},
       eprint = {1608.04994},
 primaryClass = {astro-ph.GA},
       adsurl = {https://ui.adsabs.harvard.edu/abs/2017MNRAS.465.1699M}
}

@ARTICLE{2020AJ....160...82S,
       author = {{Steinmetz}, Matthias and {Matijevi{\v{c}}}, Gal and {Enke}, Harry and {Zwitter}, Toma{\v{z}} and {Guiglion}, Guillaume and {McMillan}, Paul J. and {Kordopatis}, Georges and {Valentini}, Marica and {Chiappini}, Cristina and {Casagrande}, Luca and {Wojno}, Jennifer and {Anguiano}, Borja and {Bienaym{\'e}}, Olivier and {Bijaoui}, Albert and {Binney}, James and {Burton}, Donna and {Cass}, Paul and {de Laverny}, Patrick and {Fiegert}, Kristin and {Freeman}, Kenneth and {Fulbright}, Jon P. and {Gibson}, Brad K. and {Gilmore}, Gerard and {Grebel}, Eva K. and {Helmi}, Amina and {Kunder}, Andrea and {Munari}, Ulisse and {Navarro}, Julio F. and {Parker}, Quentin and {Ruchti}, Gregory R. and {Recio-Blanco}, Alejandra and {Reid}, Warren and {Seabroke}, George M. and {Siviero}, Alessandro and {Siebert}, Arnaud and {Stupar}, Milorad and {Watson}, Fred and {Williams}, Mary E.~K. and {Wyse}, Rosemary F.~G. and {Anders}, Friedrich and {Antoja}, Teresa and {Birko}, Danijela and {Bland-Hawthorn}, Joss and {Bossini}, Diego and {Garc{\'\i}a}, Rafael A. and {Carrillo}, Ismael and {Chaplin}, William J. and {Elsworth}, Yvonne and {Famaey}, Benoit and {Gerhard}, Ortwin and {Jofre}, Paula and {Just}, Andreas and {Mathur}, Savita and {Miglio}, Andrea and {Minchev}, Ivan and {Monari}, Giacomo and {Mosser}, Benoit and {Ritter}, Andreas and {Rodrigues}, Thaise S. and {Scholz}, Ralf-Dieter and {Sharma}, Sanjib and {Sysoliatina}, Kseniia and {RAVE Collaboration}},
        title = "{The Sixth Data Release of the Radial Velocity Experiment (RAVE). I. Survey Description, Spectra, and Radial Velocities}",
      journal = {\aj},
         year = 2020,
        month = aug,
       volume = {160},
       number = {2},
          eid = {82},
        pages = {82},
          doi = {10.3847/1538-3881/ab9ab9},
archivePrefix = {arXiv},
       eprint = {2002.04377},
 primaryClass = {astro-ph.SR},
       adsurl = {https://ui.adsabs.harvard.edu/abs/2020AJ....160...82S}
}

@INPROCEEDINGS{2003ASPC..303...87S,
       author = {{Shugarov}, S. and {Pavlenko}, E. and {Malanushenko}, V.},
        title = "{The Nature of a Twelve-Year Periodicity in the Symbiotic Nova RT Ser}",
    booktitle = {Symbiotic Stars Probing Stellar Evolution},
         year = 2003,
       editor = {{Corradi}, R.~L.~M. and {Mikolajewska}, J. and {Mahoney}, T.~J.},
       series = {Astronomical Society of the Pacific Conference Series},
       volume = {303},
        month = jan,
        pages = {87-91},
       adsurl = {https://ui.adsabs.harvard.edu/abs/2003ASPC..303...87S}
}

@ARTICLE{2024A&A...682A...7B,
       author = {{Beck}, P.~G. and {Grossmann}, D.~H. and {Steinwender}, L. and {Schimak}, L.~S. and {Muntean}, N. and {Vrard}, M. and {Patton}, R.~A. and {Merc}, J. and {Mathur}, S. and {Garcia}, R.~A. and {Pinsonneault}, M.~H. and {Rowan}, D.~M. and {Gaulme}, P. and {Allende Prieto}, C. and {Arellano-C{\'o}rdova}, K.~Z. and {Cao}, L. and {Corsaro}, E. and {Creevey}, O. and {Hambleton}, K.~M. and {Hanslmeier}, A. and {Holl}, B. and {Johnson}, J. and {Mathis}, S. and {Godoy-Rivera}, D. and {S{\'\i}mon-D{\'\i}az}, S. and {Zinn}, J.~C.},
        title = "{Constraining stellar and orbital co-evolution through ensemble seismology of solar-like oscillators in binary systems. A census of oscillating red giants and dwarf stars in Gaia DR3 binaries}",
      journal = {\aap},
         year = 2024,
        month = feb,
       volume = {682},
          eid = {A7},
        pages = {A7},
          doi = {10.1051/0004-6361/202346810},
archivePrefix = {arXiv},
       eprint = {2307.10812},
 primaryClass = {astro-ph.SR},
       adsurl = {https://ui.adsabs.harvard.edu/abs/2024A&A...682A...7B}
}

@ARTICLE{1997ApJ...489..254C,
       author = {{Chakrabarty}, Deepto and {Roche}, Paul},
        title = "{The Symbiotic Neutron Star Binary GX 1+4/V2116 Ophiuchi}",
      journal = {\apj},
         year = 1997,
        month = nov,
       volume = {489},
       number = {1},
        pages = {254-271},
          doi = {10.1086/304779},
archivePrefix = {arXiv},
       eprint = {astro-ph/9706048},
 primaryClass = {astro-ph},
       adsurl = {https://ui.adsabs.harvard.edu/abs/1997ApJ...489..254C}
}

@ARTICLE{2016MNRAS.461L...1M,
       author = {{Mukai}, K. and {Luna}, G.~J.~M. and {Cusumano}, G. and {Segreto}, A. and {Munari}, U. and {Sokoloski}, J.~L. and {Lucy}, A.~B. and {Nelson}, T. and {Nu{\~n}ez}, N.~E.},
        title = "{SU Lyncis, a hard X-ray bright M giant: clues point to a large hidden population of symbiotic stars}",
      journal = {\mnras},
         year = 2016,
        month = sep,
       volume = {461},
       number = {1},
        pages = {L1-L5},
          doi = {10.1093/mnrasl/slw087},
archivePrefix = {arXiv},
       eprint = {1604.08483},
 primaryClass = {astro-ph.SR},
       adsurl = {https://ui.adsabs.harvard.edu/abs/2016MNRAS.461L...1M}
}

@ARTICLE{2022ApJS..259...35A,
       author = {{Abdurro'uf} and {Accetta}, Katherine and {Aerts}, Conny and {Silva Aguirre}, V{\'\i}ctor and {Ahumada}, Romina and {Ajgaonkar}, Nikhil and {Filiz Ak}, N. and {Alam}, Shadab and {Allende Prieto}, Carlos and {Almeida}, Andr{\'e}s and {Anders}, Friedrich and {Anderson}, Scott F. and {Andrews}, Brett H. and {Anguiano}, Borja and {Aquino-Ort{\'\i}z}, Erik and {Arag{\'o}n-Salamanca}, Alfonso and {Argudo-Fern{\'a}ndez}, Maria and {Ata}, Metin and {Aubert}, Marie and {Avila-Reese}, Vladimir and {Badenes}, Carles and {Barb{\'a}}, Rodolfo H. and {Barger}, Kat and {Barrera-Ballesteros}, Jorge K. and {Beaton}, Rachael L. and {Beers}, Timothy C. and {Belfiore}, Francesco and {Bender}, Chad F. and {Bernardi}, Mariangela and {Bershady}, Matthew A. and {Beutler}, Florian and {Bidin}, Christian Moni and {Bird}, Jonathan C. and {Bizyaev}, Dmitry and {Blanc}, Guillermo A. and {Blanton}, Michael R. and {Boardman}, Nicholas Fraser and {Bolton}, Adam S. and {Boquien}, M{\'e}d{\'e}ric and {Borissova}, Jura and {Bovy}, Jo and {Brandt}, W.~N. and {Brown}, Jordan and {Brownstein}, Joel R. and {Brusa}, Marcella and {Buchner}, Johannes and {Bundy}, Kevin and {Burchett}, Joseph N. and {Bureau}, Martin and {Burgasser}, Adam and {Cabang}, Tuesday K. and {Campbell}, Stephanie and {Cappellari}, Michele and {Carlberg}, Joleen K. and {Wanderley}, F{\'a}bio Carneiro and {Carrera}, Ricardo and {Cash}, Jennifer and {Chen}, Yan-Ping and {Chen}, Wei-Huai and {Cherinka}, Brian and {Chiappini}, Cristina and {Choi}, Peter Doohyun and {Chojnowski}, S. Drew and {Chung}, Haeun and {Clerc}, Nicolas and {Cohen}, Roger E. and {Comerford}, Julia M. and {Comparat}, Johan and {da Costa}, Luiz and {Covey}, Kevin and {Crane}, Jeffrey D. and {Cruz-Gonzalez}, Irene and {Culhane}, Connor and {Cunha}, Katia and {Dai}, Y. Sophia and {Damke}, Guillermo and {Darling}, Jeremy and {Davidson}, Jr., James W. and {Davies}, Roger and {Dawson}, Kyle and {De Lee}, Nathan and {Diamond-Stanic}, Aleksandar M. and {Cano-D{\'\i}az}, Mariana and {S{\'a}nchez}, Helena Dom{\'\i}nguez and {Donor}, John and {Duckworth}, Chris and {Dwelly}, Tom and {Eisenstein}, Daniel J. and {Elsworth}, Yvonne P. and {Emsellem}, Eric and {Eracleous}, Mike and {Escoffier}, Stephanie and {Fan}, Xiaohui and {Farr}, Emily and {Feng}, Shuai and {Fern{\'a}ndez-Trincado}, Jos{\'e} G. and {Feuillet}, Diane and {Filipp}, Andreas and {Fillingham}, Sean P. and {Frinchaboy}, Peter M. and {Fromenteau}, Sebastien and {Galbany}, Llu{\'\i}s and {Garc{\'\i}a}, Rafael A. and {Garc{\'\i}a-Hern{\'a}ndez}, D.~A. and {Ge}, Junqiang and {Geisler}, Doug and {Gelfand}, Joseph and {G{\'e}ron}, Tobias and {Gibson}, Benjamin J. and {Goddy}, Julian and {Godoy-Rivera}, Diego and {Grabowski}, Kathleen and {Green}, Paul J. and {Greener}, Michael and {Grier}, Catherine J. and {Griffith}, Emily and {Guo}, Hong and {Guy}, Julien and {Hadjara}, Massinissa and {Harding}, Paul and {Hasselquist}, Sten and {Hayes}, Christian R. and {Hearty}, Fred and {Hern{\'a}ndez}, Jes{\'u}s and {Hill}, Lewis and {Hogg}, David W. and {Holtzman}, Jon A. and {Horta}, Danny and {Hsieh}, Bau-Ching and {Hsu}, Chin-Hao and {Hsu}, Yun-Hsin and {Huber}, Daniel and {Huertas-Company}, Marc and {Hutchinson}, Brian and {Hwang}, Ho Seong and {Ibarra-Medel}, H{\'e}ctor J. and {Chitham}, Jacob Ider and {Ilha}, Gabriele S. and {Imig}, Julie and {Jaekle}, Will and {Jayasinghe}, Tharindu and {Ji}, Xihan and {Johnson}, Jennifer A. and {Jones}, Amy and {J{\"o}nsson}, Henrik and {Katkov}, Ivan and {Khalatyan}, Dr., Arman and {Kinemuchi}, Karen and {Kisku}, Shobhit and {Knapen}, Johan H. and {Kneib}, Jean-Paul and {Kollmeier}, Juna A. and {Kong}, Miranda and {Kounkel}, Marina and {Kreckel}, Kathryn and {Krishnarao}, Dhanesh and {Lacerna}, Ivan and {Lane}, Richard R. and {Langgin}, Rachel and {Lavender}, Ramon and {Law}, David R. and {Lazarz}, Daniel and {Leung}, Henry W. and {Leung}, Ho-Hin and {Lewis}, Hannah M. and {Li}, Cheng and {Li}, Ran and {Lian}, Jianhui and {Liang}, Fu-Heng and {Lin}, Lihwai and {Lin}, Yen-Ting and {Lin}, Sicheng and {Lintott}, Chris and {Long}, Dan and {Longa-Pe{\~n}a}, Pen{\'e}lope and {L{\'o}pez-Cob{\'a}}, Carlos and {Lu}, Shengdong and {Lundgren}, Britt F. and {Luo}, Yuanze and {Mackereth}, J. Ted and {de la Macorra}, Axel and {Mahadevan}, Suvrath and {Majewski}, Steven R. and {Manchado}, Arturo and {Mandeville}, Travis and {Maraston}, Claudia and {Margalef-Bentabol}, Berta and {Masseron}, Thomas and {Masters}, Karen L. and {Mathur}, Savita and {McDermid}, Richard M. and {Mckay}, Myles and {Merloni}, Andrea and {Merrifield}, Michael and {Meszaros}, Szabolcs and {Miglio}, Andrea and {Di Mille}, Francesco and {Minniti}, Dante and {Minsley}, Rebecca and {Monachesi}, Antonela},
        title = "{The Seventeenth Data Release of the Sloan Digital Sky Surveys: Complete Release of MaNGA, MaStar, and APOGEE-2 Data}",
      journal = {\apjs},
         year = 2022,
        month = apr,
       volume = {259},
       number = {2},
          eid = {35},
        pages = {35},
          doi = {10.3847/1538-4365/ac4414},
archivePrefix = {arXiv},
       eprint = {2112.02026},
 primaryClass = {astro-ph.GA},
       adsurl = {https://ui.adsabs.harvard.edu/abs/2022ApJS..259...35A}
}

@ARTICLE{2025A&A...698A.155L,
       author = {{Laversveiler}, Marco and {Gon{\c{c}}alves}, Denise R. and {Rocha-Pinto}, Helio J. and {Merc}, Jaroslav},
        title = "{The local group symbiotic star population and its tenuous link with type Ia supernovae}",
      journal = {\aap},
         year = 2025,
        month = jun,
       volume = {698},
          eid = {A155},
        pages = {A155},
          doi = {10.1051/0004-6361/202451548},
archivePrefix = {arXiv},
       eprint = {2504.02090},
 primaryClass = {astro-ph.GA},
       adsurl = {https://ui.adsabs.harvard.edu/abs/2025A&A...698A.155L}
}

@ARTICLE{2013A&A...559A...6L,
       author = {{Luna}, G.~J.~M. and {Sokoloski}, J.~L. and {Mukai}, K. and {Nelson}, T.},
        title = "{Symbiotic stars in X-rays}",
      journal = {\aap},
         year = 2013,
        month = nov,
       volume = {559},
          eid = {A6},
        pages = {A6},
          doi = {10.1051/0004-6361/201220792},
archivePrefix = {arXiv},
       eprint = {1211.6082},
 primaryClass = {astro-ph.SR},
       adsurl = {https://ui.adsabs.harvard.edu/abs/2013A&A...559A...6L}
}

@ARTICLE{2024A&A...690A.243S,
       author = {{Schwope}, A.~D. and {Knauff}, K. and {Kurpas}, J. and {Salvato}, M. and {Stelzer}, B. and {St{\"u}tz}, L. and {Tub{\'\i}n-Arenas}, D.},
        title = "{A first systematic characterization of cataclysmic variables in SRG/eROSITA surveys}",
      journal = {\aap},
         year = 2024,
        month = oct,
       volume = {690},
          eid = {A243},
        pages = {A243},
          doi = {10.1051/0004-6361/202450537},
archivePrefix = {arXiv},
       eprint = {2407.20903},
 primaryClass = {astro-ph.SR},
       adsurl = {https://ui.adsabs.harvard.edu/abs/2024A&A...690A.243S}
}

@ARTICLE{2023A&A...674A..14R,
       author = {{Rimoldini}, Lorenzo and {Holl}, Berry and {Gavras}, Panagiotis and {Audard}, Marc and {De Ridder}, Joris and {Mowlavi}, Nami and {Nienartowicz}, Krzysztof and {Jevardat de Fombelle}, Gr{\'e}gory and {Lecoeur-Ta{\"\i}bi}, Isabelle and {Karbevska}, Lea and {Evans}, Dafydd W. and {{\'A}brah{\'a}m}, P{\'e}ter and {Carnerero}, Maria I. and {Clementini}, Gisella and {Distefano}, Elisa and {Garofalo}, Alessia and {Garc{\'\i}a-Lario}, Pedro and {Gomel}, Roy and {Klioner}, Sergei A. and {Kruszy{\'n}ska}, Katarzyna and {Lanzafame}, Alessandro C. and {Lebzelter}, Thomas and {Marton}, G{\'a}bor and {Mazeh}, Tsevi and {Molinaro}, Roberto and {Panahi}, Aviad and {Raiteri}, Claudia M. and {Ripepi}, Vincenzo and {Szabados}, L{\'a}szl{\'o} and {Teyssier}, David and {Trabucchi}, Michele and {Wyrzykowski}, {\L}ukasz and {Zucker}, Shay and {Eyer}, Laurent},
        title = "{Gaia Data Release 3. All-sky classification of 12.4 million variable sources into 25 classes}",
      journal = {\aap},
         year = 2023,
        month = jun,
       volume = {674},
          eid = {A14},
        pages = {A14},
          doi = {10.1051/0004-6361/202245591},
archivePrefix = {arXiv},
       eprint = {2211.17238},
 primaryClass = {astro-ph.GA},
       adsurl = {https://ui.adsabs.harvard.edu/abs/2023A&A...674A..14R}
}

@ARTICLE{2000A&AS..146..407B,
       author = {{Belczy{\'n}ski}, K. and {Miko{\l}ajewska}, J. and {Munari}, U. and {Ivison}, R.~J. and {Friedjung}, M.},
        title = "{A catalogue of symbiotic stars}",
      journal = {\aaps},
         year = 2000,
        month = nov,
       volume = {146},
        pages = {407-435},
          doi = {10.1051/aas:2000280},
archivePrefix = {arXiv},
       eprint = {astro-ph/0005547},
 primaryClass = {astro-ph},
       adsurl = {https://ui.adsabs.harvard.edu/abs/2000A&AS..146..407B}
}

@ARTICLE{2019ApJS..240...21A,
       author = {{Akras}, Stavros and {Guzman-Ramirez}, Lizette and {Leal-Ferreira}, Marcelo L. and {Ramos-Larios}, Gerardo},
        title = "{A Census of Symbiotic Stars in the 2MASS, WISE, and Gaia Surveys}",
      journal = {\apjs},
         year = 2019,
        month = feb,
       volume = {240},
       number = {2},
          eid = {21},
        pages = {21},
          doi = {10.3847/1538-4365/aaf88c},
archivePrefix = {arXiv},
       eprint = {1902.01451},
 primaryClass = {astro-ph.SR},
       adsurl = {https://ui.adsabs.harvard.edu/abs/2019ApJS..240...21A}
}

@ARTICLE{1999MNRAS.305..190M,
       author = {{Miko{\l}ajewska}, Joanna and {Brandi}, Estela and {Hack}, Warren and {Whitelock}, Patricia A. and {Barba}, Rodolfo and {Garcia}, Lia and {Marang}, Freddy},
        title = "{The symbiotic binary system RX Puppis: a possible recurrent nova with a Mira companion}",
      journal = {\mnras},
         year = 1999,
        month = may,
       volume = {305},
       number = {1},
        pages = {190-210},
          doi = {10.1046/j.1365-8711.1999.02449.x},
archivePrefix = {arXiv},
       eprint = {astro-ph/9901044},
 primaryClass = {astro-ph},
       adsurl = {https://ui.adsabs.harvard.edu/abs/1999MNRAS.305..190M}
}

@ARTICLE{2023A&A...674A...1G,
       author = {{Gaia Collaboration} and {Vallenari}, A. and {Brown}, A.~G.~A. and {Prusti}, T. and {de Bruijne}, J.~H.~J. and {Arenou}, F. and {Babusiaux}, C. and {Biermann}, M. and {Creevey}, O.~L. and {Ducourant}, C. and {Evans}, D.~W. and {Eyer}, L. and {Guerra}, R. and {Hutton}, A. and {Jordi}, C. and {Klioner}, S.~A. and {Lammers}, U.~L. and {Lindegren}, L. and {Luri}, X. and {Mignard}, F. and {Panem}, C. and {Pourbaix}, D. and {Randich}, S. and {Sartoretti}, P. and {Soubiran}, C. and {Tanga}, P. and {Walton}, N.~A. and {Bailer-Jones}, C.~A.~L. and {Bastian}, U. and {Drimmel}, R. and {Jansen}, F. and {Katz}, D. and {Lattanzi}, M.~G. and {van Leeuwen}, F. and {Bakker}, J. and {Cacciari}, C. and {Casta{\~n}eda}, J. and {De Angeli}, F. and {Fabricius}, C. and {Fouesneau}, M. and {Fr{\'e}mat}, Y. and {Galluccio}, L. and {Guerrier}, A. and {Heiter}, U. and {Masana}, E. and {Messineo}, R. and {Mowlavi}, N. and {Nicolas}, C. and {Nienartowicz}, K. and {Pailler}, F. and {Panuzzo}, P. and {Riclet}, F. and {Roux}, W. and {Seabroke}, G.~M. and {Sordo}, R. and {Th{\'e}venin}, F. and {Gracia-Abril}, G. and {Portell}, J. and {Teyssier}, D. and {Altmann}, M. and {Andrae}, R. and {Audard}, M. and {Bellas-Velidis}, I. and {Benson}, K. and {Berthier}, J. and {Blomme}, R. and {Burgess}, P.~W. and {Busonero}, D. and {Busso}, G. and {C{\'a}novas}, H. and {Carry}, B. and {Cellino}, A. and {Cheek}, N. and {Clementini}, G. and {Damerdji}, Y. and {Davidson}, M. and {de Teodoro}, P. and {Nu{\~n}ez Campos}, M. and {Delchambre}, L. and {Dell'Oro}, A. and {Esquej}, P. and {Fern{\'a}ndez-Hern{\'a}ndez}, J. and {Fraile}, E. and {Garabato}, D. and {Garc{\'\i}a-Lario}, P. and {Gosset}, E. and {Haigron}, R. and {Halbwachs}, J. -L. and {Hambly}, N.~C. and {Harrison}, D.~L. and {Hern{\'a}ndez}, J. and {Hestroffer}, D. and {Hodgkin}, S.~T. and {Holl}, B. and {Jan{\ss}en}, K. and {Jevardat de Fombelle}, G. and {Jordan}, S. and {Krone-Martins}, A. and {Lanzafame}, A.~C. and {L{\"o}ffler}, W. and {Marchal}, O. and {Marrese}, P.~M. and {Moitinho}, A. and {Muinonen}, K. and {Osborne}, P. and {Pancino}, E. and {Pauwels}, T. and {Recio-Blanco}, A. and {Reyl{\'e}}, C. and {Riello}, M. and {Rimoldini}, L. and {Roegiers}, T. and {Rybizki}, J. and {Sarro}, L.~M. and {Siopis}, C. and {Smith}, M. and {Sozzetti}, A. and {Utrilla}, E. and {van Leeuwen}, M. and {Abbas}, U. and {{\'A}brah{\'a}m}, P. and {Abreu Aramburu}, A. and {Aerts}, C. and {Aguado}, J.~J. and {Ajaj}, M. and {Aldea-Montero}, F. and {Altavilla}, G. and {{\'A}lvarez}, M.~A. and {Alves}, J. and {Anders}, F. and {Anderson}, R.~I. and {Anglada Varela}, E. and {Antoja}, T. and {Baines}, D. and {Baker}, S.~G. and {Balaguer-N{\'u}{\~n}ez}, L. and {Balbinot}, E. and {Balog}, Z. and {Barache}, C. and {Barbato}, D. and {Barros}, M. and {Barstow}, M.~A. and {Bartolom{\'e}}, S. and {Bassilana}, J. -L. and {Bauchet}, N. and {Becciani}, U. and {Bellazzini}, M. and {Berihuete}, A. and {Bernet}, M. and {Bertone}, S. and {Bianchi}, L. and {Binnenfeld}, A. and {Blanco-Cuaresma}, S. and {Blazere}, A. and {Boch}, T. and {Bombrun}, A. and {Bossini}, D. and {Bouquillon}, S. and {Bragaglia}, A. and {Bramante}, L. and {Breedt}, E. and {Bressan}, A. and {Brouillet}, N. and {Brugaletta}, E. and {Bucciarelli}, B. and {Burlacu}, A. and {Butkevich}, A.~G. and {Buzzi}, R. and {Caffau}, E. and {Cancelliere}, R. and {Cantat-Gaudin}, T. and {Carballo}, R. and {Carlucci}, T. and {Carnerero}, M.~I. and {Carrasco}, J.~M. and {Casamiquela}, L. and {Castellani}, M. and {Castro-Ginard}, A. and {Chaoul}, L. and {Charlot}, P. and {Chemin}, L. and {Chiaramida}, V. and {Chiavassa}, A. and {Chornay}, N. and {Comoretto}, G. and {Contursi}, G. and {Cooper}, W.~J. and {Cornez}, T. and {Cowell}, S. and {Crifo}, F. and {Cropper}, M. and {Crosta}, M. and {Crowley}, C. and {Dafonte}, C. and {Dapergolas}, A. and {David}, M. and {David}, P. and {de Laverny}, P. and {De Luise}, F. and {De March}, R.},
        title = "{Gaia Data Release 3. Summary of the content and survey properties}",
      journal = {\aap},
         year = 2023,
        month = jun,
       volume = {674},
          eid = {A1},
        pages = {A1},
          doi = {10.1051/0004-6361/202243940},
archivePrefix = {arXiv},
       eprint = {2208.00211},
 primaryClass = {astro-ph.GA},
       adsurl = {https://ui.adsabs.harvard.edu/abs/2023A&A...674A...1G}
}

@ARTICLE{2016A&A...595A...1G,
       author = {{Gaia Collaboration} and {Prusti}, T. and {de Bruijne}, J.~H.~J. and {Brown}, A.~G.~A. and {Vallenari}, A. and {Babusiaux}, C. and {Bailer-Jones}, C.~A.~L. and {Bastian}, U. and {Biermann}, M. and {Evans}, D.~W. and {Eyer}, L. and {Jansen}, F. and {Jordi}, C. and {Klioner}, S.~A. and {Lammers}, U. and {Lindegren}, L. and {Luri}, X. and {Mignard}, F. and {Milligan}, D.~J. and {Panem}, C. and {Poinsignon}, V. and {Pourbaix}, D. and {Randich}, S. and {Sarri}, G. and {Sartoretti}, P. and {Siddiqui}, H.~I. and {Soubiran}, C. and {Valette}, V. and {van Leeuwen}, F. and {Walton}, N.~A. and {Aerts}, C. and {Arenou}, F. and {Cropper}, M. and {Drimmel}, R. and {H{\o}g}, E. and {Katz}, D. and {Lattanzi}, M.~G. and {O'Mullane}, W. and {Grebel}, E.~K. and {Holland}, A.~D. and {Huc}, C. and {Passot}, X. and {Bramante}, L. and {Cacciari}, C. and {Casta{\~n}eda}, J. and {Chaoul}, L. and {Cheek}, N. and {De Angeli}, F. and {Fabricius}, C. and {Guerra}, R. and {Hern{\'a}ndez}, J. and {Jean-Antoine-Piccolo}, A. and {Masana}, E. and {Messineo}, R. and {Mowlavi}, N. and {Nienartowicz}, K. and {Ord{\'o}{\~n}ez-Blanco}, D. and {Panuzzo}, P. and {Portell}, J. and {Richards}, P.~J. and {Riello}, M. and {Seabroke}, G.~M. and {Tanga}, P. and {Th{\'e}venin}, F. and {Torra}, J. and {Els}, S.~G. and {Gracia-Abril}, G. and {Comoretto}, G. and {Garcia-Reinaldos}, M. and {Lock}, T. and {Mercier}, E. and {Altmann}, M. and {Andrae}, R. and {Astraatmadja}, T.~L. and {Bellas-Velidis}, I. and {Benson}, K. and {Berthier}, J. and {Blomme}, R. and {Busso}, G. and {Carry}, B. and {Cellino}, A. and {Clementini}, G. and {Cowell}, S. and {Creevey}, O. and {Cuypers}, J. and {Davidson}, M. and {De Ridder}, J. and {de Torres}, A. and {Delchambre}, L. and {Dell'Oro}, A. and {Ducourant}, C. and {Fr{\'e}mat}, Y. and {Garc{\'\i}a-Torres}, M. and {Gosset}, E. and {Halbwachs}, J. -L. and {Hambly}, N.~C. and {Harrison}, D.~L. and {Hauser}, M. and {Hestroffer}, D. and {Hodgkin}, S.~T. and {Huckle}, H.~E. and {Hutton}, A. and {Jasniewicz}, G. and {Jordan}, S. and {Kontizas}, M. and {Korn}, A.~J. and {Lanzafame}, A.~C. and {Manteiga}, M. and {Moitinho}, A. and {Muinonen}, K. and {Osinde}, J. and {Pancino}, E. and {Pauwels}, T. and {Petit}, J. -M. and {Recio-Blanco}, A. and {Robin}, A.~C. and {Sarro}, L.~M. and {Siopis}, C. and {Smith}, M. and {Smith}, K.~W. and {Sozzetti}, A. and {Thuillot}, W. and {van Reeven}, W. and {Viala}, Y. and {Abbas}, U. and {Abreu Aramburu}, A. and {Accart}, S. and {Aguado}, J.~J. and {Allan}, P.~M. and {Allasia}, W. and {Altavilla}, G. and {{\'A}lvarez}, M.~A. and {Alves}, J. and {Anderson}, R.~I. and {Andrei}, A.~H. and {Anglada Varela}, E. and {Antiche}, E. and {Antoja}, T. and {Ant{\'o}n}, S. and {Arcay}, B. and {Atzei}, A. and {Ayache}, L. and {Bach}, N. and {Baker}, S.~G. and {Balaguer-N{\'u}{\~n}ez}, L. and {Barache}, C. and {Barata}, C. and {Barbier}, A. and {Barblan}, F. and {Baroni}, M. and {Barrado y Navascu{\'e}s}, D. and {Barros}, M. and {Barstow}, M.~A. and {Becciani}, U. and {Bellazzini}, M. and {Bellei}, G. and {Bello Garc{\'\i}a}, A. and {Belokurov}, V. and {Bendjoya}, P. and {Berihuete}, A. and {Bianchi}, L. and {Bienaym{\'e}}, O. and {Billebaud}, F. and {Blagorodnova}, N. and {Blanco-Cuaresma}, S. and {Boch}, T. and {Bombrun}, A. and {Borrachero}, R. and {Bouquillon}, S. and {Bourda}, G. and {Bouy}, H. and {Bragaglia}, A. and {Breddels}, M.~A. and {Brouillet}, N. and {Br{\"u}semeister}, T. and {Bucciarelli}, B. and {Budnik}, F. and {Burgess}, P. and {Burgon}, R. and {Burlacu}, A. and {Busonero}, D. and {Buzzi}, R. and {Caffau}, E. and {Cambras}, J. and {Campbell}, H. and {Cancelliere}, R. and {Cantat-Gaudin}, T. and {Carlucci}, T. and {Carrasco}, J.~M. and {Castellani}, M. and {Charlot}, P. and {Charnas}, J. and {Charvet}, P. and {Chassat}, F. and {Chiavassa}, A. and {Clotet}, M. and {Cocozza}, G. and {Collins}, R.~S. and {Collins}, P. and {Costigan}, G.},
        title = "{The Gaia mission}",
      journal = {\aap},
         year = 2016,
        month = nov,
       volume = {595},
          eid = {A1},
        pages = {A1},
          doi = {10.1051/0004-6361/201629272},
archivePrefix = {arXiv},
       eprint = {1609.04153},
 primaryClass = {astro-ph.IM},
       adsurl = {https://ui.adsabs.harvard.edu/abs/2016A&A...595A...1G}
}

@ARTICLE{2023A&A...674A...2D,
       author = {{De Angeli}, F. and {Weiler}, M. and {Montegriffo}, P. and {Evans}, D.~W. and {Riello}, M. and {Andrae}, R. and {Carrasco}, J.~M. and {Busso}, G. and {Burgess}, P.~W. and {Cacciari}, C. and {Davidson}, M. and {Harrison}, D.~L. and {Hodgkin}, S.~T. and {Jordi}, C. and {Osborne}, P.~J. and {Pancino}, E. and {Altavilla}, G. and {Barstow}, M.~A. and {Bailer-Jones}, C.~A.~L. and {Bellazzini}, M. and {Brown}, A.~G.~A. and {Castellani}, M. and {Cowell}, S. and {Delchambre}, L. and {De Luise}, F. and {Diener}, C. and {Fabricius}, C. and {Fouesneau}, M. and {Fr{\'e}mat}, Y. and {Gilmore}, G. and {Giuffrida}, G. and {Hambly}, N.~C. and {Hidalgo}, S. and {Holland}, G. and {Kostrzewa-Rutkowska}, Z. and {van Leeuwen}, F. and {Lobel}, A. and {Marinoni}, S. and {Miller}, N. and {Pagani}, C. and {Palaversa}, L. and {Piersimoni}, A.~M. and {Pulone}, L. and {Ragaini}, S. and {Rainer}, M. and {Richards}, P.~J. and {Rixon}, G.~T. and {Ruz-Mieres}, D. and {Sanna}, N. and {Sarro}, L.~M. and {Rowell}, N. and {Sordo}, R. and {Walton}, N.~A. and {Yoldas}, A.},
        title = "{Gaia Data Release 3. Processing and validation of BP/RP low-resolution spectral data}",
      journal = {\aap},
         year = 2023,
        month = jun,
       volume = {674},
          eid = {A2},
        pages = {A2},
          doi = {10.1051/0004-6361/202243680},
archivePrefix = {arXiv},
       eprint = {2206.06143},
 primaryClass = {astro-ph.IM},
       adsurl = {https://ui.adsabs.harvard.edu/abs/2023A&A...674A...2D}
}

@ARTICLE{2023A&A...674A...3M,
       author = {{Montegriffo}, P. and {De Angeli}, F. and {Andrae}, R. and {Riello}, M. and {Pancino}, E. and {Sanna}, N. and {Bellazzini}, M. and {Evans}, D.~W. and {Carrasco}, J.~M. and {Sordo}, R. and {Busso}, G. and {Cacciari}, C. and {Jordi}, C. and {van Leeuwen}, F. and {Vallenari}, A. and {Altavilla}, G. and {Barstow}, M.~A. and {Brown}, A.~G.~A. and {Burgess}, P.~W. and {Castellani}, M. and {Cowell}, S. and {Davidson}, M. and {De Luise}, F. and {Delchambre}, L. and {Diener}, C. and {Fabricius}, C. and {Fr{\'e}mat}, Y. and {Fouesneau}, M. and {Gilmore}, G. and {Giuffrida}, G. and {Hambly}, N.~C. and {Harrison}, D.~L. and {Hidalgo}, S. and {Hodgkin}, S.~T. and {Holland}, G. and {Marinoni}, S. and {Osborne}, P.~J. and {Pagani}, C. and {Palaversa}, L. and {Piersimoni}, A.~M. and {Pulone}, L. and {Ragaini}, S. and {Rainer}, M. and {Richards}, P.~J. and {Rowell}, N. and {Ruz-Mieres}, D. and {Sarro}, L.~M. and {Walton}, N.~A. and {Yoldas}, A.},
        title = "{Gaia Data Release 3. External calibration of BP/RP low-resolution spectroscopic data}",
      journal = {\aap},
         year = 2023,
        month = jun,
       volume = {674},
          eid = {A3},
        pages = {A3},
          doi = {10.1051/0004-6361/202243880},
archivePrefix = {arXiv},
       eprint = {2206.06205},
 primaryClass = {astro-ph.IM},
       adsurl = {https://ui.adsabs.harvard.edu/abs/2023A&A...674A...3M}
}

@ARTICLE{2023A&A...674A...4E,
       author = {{Evans}, D.~W. and {Eyer}, L. and {Busso}, G. and {Riello}, M. and {De Angeli}, F. and {Burgess}, P.~W. and {Audard}, M. and {Clementini}, G. and {Garofalo}, A. and {Holl}, B. and {Jevardat de Fombelle}, G. and {Lanzafame}, A.~C. and {Lecoeur-Taibi}, I. and {Mowlavi}, N. and {Nienartowicz}, K. and {Palaversa}, L. and {Rimoldini}, L.},
        title = "{Gaia Data Release 3. The Gaia Andromeda Photometric Survey}",
      journal = {\aap},
         year = 2023,
        month = jun,
       volume = {674},
          eid = {A4},
        pages = {A4},
          doi = {10.1051/0004-6361/202244204},
archivePrefix = {arXiv},
       eprint = {2206.05591},
 primaryClass = {astro-ph.IM},
       adsurl = {https://ui.adsabs.harvard.edu/abs/2023A&A...674A...4E}
}

@ARTICLE{2023A&A...674A...5K,
       author = {{Katz}, D. and {Sartoretti}, P. and {Guerrier}, A. and {Panuzzo}, P. and {Seabroke}, G.~M. and {Th{\'e}venin}, F. and {Cropper}, M. and {Benson}, K. and {Blomme}, R. and {Haigron}, R. and {Marchal}, O. and {Smith}, M. and {Baker}, S. and {Chemin}, L. and {Damerdji}, Y. and {David}, M. and {Dolding}, C. and {Fr{\'e}mat}, Y. and {Gosset}, E. and {Jan{\ss}en}, K. and {Jasniewicz}, G. and {Lobel}, A. and {Plum}, G. and {Samaras}, N. and {Snaith}, O. and {Soubiran}, C. and {Vanel}, O. and {Zwitter}, T. and {Antoja}, T. and {Arenou}, F. and {Babusiaux}, C. and {Brouillet}, N. and {Caffau}, E. and {Di Matteo}, P. and {Fabre}, C. and {Fabricius}, C. and {Fragkoudi}, F. and {Haywood}, M. and {Huckle}, H.~E. and {Hottier}, C. and {Lasne}, Y. and {Leclerc}, N. and {Mastrobuono-Battisti}, A. and {Royer}, F. and {Teyssier}, D. and {Zorec}, J. and {Crifo}, F. and {Jean-Antoine Piccolo}, A. and {Turon}, C. and {Viala}, Y.},
        title = "{Gaia Data Release 3. Properties and validation of the radial velocities}",
      journal = {\aap},
         year = 2023,
        month = jun,
       volume = {674},
          eid = {A5},
        pages = {A5},
          doi = {10.1051/0004-6361/202244220},
archivePrefix = {arXiv},
       eprint = {2206.05902},
 primaryClass = {astro-ph.GA},
       adsurl = {https://ui.adsabs.harvard.edu/abs/2023A&A...674A...5K}
}

@ARTICLE{2025A&A...693A.124G,
       author = {{Gosset}, E. and {Damerdji}, Y. and {Morel}, T. and {Delchambre}, L. and {Halbwachs}, J. -L. and {Sadowski}, G. and {Pourbaix}, D. and {Sozzetti}, A. and {Panuzzo}, P. and {Arenou}, F.},
        title = "{Gaia Data Release 3: Spectroscopic binary-star orbital solutions: The SB1 processing chain}",
      journal = {\aap},
         year = 2025,
        month = jan,
       volume = {693},
          eid = {A124},
        pages = {A124},
          doi = {10.1051/0004-6361/202450600},
archivePrefix = {arXiv},
       eprint = {2410.14372},
 primaryClass = {astro-ph.SR},
       adsurl = {https://ui.adsabs.harvard.edu/abs/2025A&A...693A.124G}
}

@ARTICLE{2023A&A...674A..13E,
       author = {{Eyer}, L. and {Audard}, M. and {Holl}, B. and {Rimoldini}, L. and {Carnerero}, M.~I. and {Clementini}, G. and {De Ridder}, J. and {Distefano}, E. and {Evans}, D.~W. and {Gavras}, P. and {Gomel}, R. and {Lebzelter}, T. and {Marton}, G. and {Mowlavi}, N. and {Panahi}, A. and {Ripepi}, V. and {Wyrzykowski}, {\L}. and {Nienartowicz}, K. and {Jevardat de Fombelle}, G. and {Lecoeur-Taibi}, I. and {Rohrbasser}, L. and {Riello}, M. and {Garc{\'\i}a-Lario}, P. and {Lanzafame}, A.~C. and {Mazeh}, T. and {Raiteri}, C.~M. and {Zucker}, S. and {{\'A}brah{\'a}m}, P. and {Aerts}, C. and {Aguado}, J.~J. and {Anderson}, R.~I. and {Bashi}, D. and {Binnenfeld}, A. and {Faigler}, S. and {Garofalo}, A. and {Karbevska}, L. and {K{\'o}sp{\'a}l}, {\'A}. and {Kruszy{\'n}ska}, K. and {Kun}, M. and {Lanza}, A.~F. and {Leccia}, S. and {Marconi}, M. and {Messina}, S. and {Molinaro}, R. and {Moln{\'a}r}, L. and {Muraveva}, T. and {Musella}, I. and {Nagy}, Z. and {Pagano}, I. and {Palaversa}, L. and {Plachy}, E. and {Pr{\v{s}}a}, A. and {Rybicki}, K.~A. and {Shahaf}, S. and {Szabados}, L. and {Szegedi-Elek}, E. and {Trabucchi}, M. and {Barblan}, F. and {Grenon}, M. and {Roelens}, M. and {S{\"u}veges}, M.},
        title = "{Gaia Data Release 3. Summary of the variability processing and analysis}",
      journal = {\aap},
         year = 2023,
        month = jun,
       volume = {674},
          eid = {A13},
        pages = {A13},
          doi = {10.1051/0004-6361/202244242},
archivePrefix = {arXiv},
       eprint = {2206.06416},
 primaryClass = {astro-ph.SR},
       adsurl = {https://ui.adsabs.harvard.edu/abs/2023A&A...674A..13E}
}

@ARTICLE{2025Galax..13...49M,
       author = {{Merc}, Jaroslav},
        title = "{Symbiotic Stars in the Era of Modern Ground- and Space-Based Surveys}",
      journal = {Galaxies},
         year = 2025,
        month = apr,
       volume = {13},
       number = {3},
          eid = {49},
        pages = {49},
          doi = {10.3390/galaxies13030049},
archivePrefix = {arXiv},
       eprint = {2504.16825},
 primaryClass = {astro-ph.SR},
       adsurl = {https://ui.adsabs.harvard.edu/abs/2025Galax..13...49M}
}

@ARTICLE{2019arXiv190901389M,
       author = {{Munari}, Ulisse},
        title = "{The Symbiotic Stars}",
      journal = {arXiv e-prints},
         year = 2019,
        month = sep,
          eid = {arXiv:1909.01389},
        pages = {arXiv:1909.01389},
          doi = {10.48550/arXiv.1909.01389},
archivePrefix = {arXiv},
       eprint = {1909.01389},
 primaryClass = {astro-ph.SR},
       adsurl = {https://ui.adsabs.harvard.edu/abs/2019arXiv190901389M}
}

@ARTICLE{2012BaltA..21....5M,
       author = {{Miko{\l}ajewska}, J.},
        title = "{Symbiotic Stars: Observations Confront Theory}",
      journal = {Baltic Astronomy},
         year = 2012,
        month = jan,
       volume = {21},
        pages = {5-12},
          doi = {10.1515/astro-2017-0352},
archivePrefix = {arXiv},
       eprint = {1110.2361},
 primaryClass = {astro-ph.SR},
       adsurl = {https://ui.adsabs.harvard.edu/abs/2012BaltA..21....5M}
}

@ARTICLE{2024A&A...688A...1C,
       author = {{Castro-Ginard}, Alfred and {Penoyre}, Zephyr and {Casey}, Andrew R. and {Brown}, Anthony G.~A. and {Belokurov}, Vasily and {Cantat-Gaudin}, Tristan and {Drimmel}, Ronald and {Fouesneau}, Morgan and {Khanna}, Shourya and {Kurbatov}, Evgeny P. and {Price-Whelan}, Adrian M. and {Rix}, Hans-Walter and {Smart}, Richard L.},
        title = "{Gaia DR3 detectability of unresolved binary systems}",
      journal = {\aap},
         year = 2024,
        month = aug,
       volume = {688},
          eid = {A1},
        pages = {A1},
          doi = {10.1051/0004-6361/202450172},
archivePrefix = {arXiv},
       eprint = {2404.14127},
 primaryClass = {astro-ph.GA},
       adsurl = {https://ui.adsabs.harvard.edu/abs/2024A&A...688A...1C}
}

@ARTICLE{2017ApJS..230...15M,
       author = {{Moe}, Maxwell and {Di Stefano}, Rosanne},
        title = "{Mind Your Ps and Qs: The Interrelation between Period (P) and Mass-ratio (Q) Distributions of Binary Stars}",
      journal = {\apjs},
         year = 2017,
        month = jun,
       volume = {230},
       number = {2},
          eid = {15},
        pages = {15},
          doi = {10.3847/1538-4365/aa6fb6},
archivePrefix = {arXiv},
       eprint = {1606.05347},
 primaryClass = {astro-ph.SR},
       adsurl = {https://ui.adsabs.harvard.edu/abs/2017ApJS..230...15M}
}

@INPROCEEDINGS{2023ASPC..534..275O,
       author = {{Offner}, S.~S.~R. and {Moe}, M. and {Kratter}, K.~M. and {Sadavoy}, S.~I. and {Jensen}, E.~L.~N. and {Tobin}, J.~J.},
        title = "{The Origin and Evolution of Multiple Star Systems}",
    booktitle = {Protostars and Planets VII},
         year = 2023,
       editor = {{Inutsuka}, S. and {Aikawa}, Y. and {Muto}, T. and {Tomida}, K. and {Tamura}, M.},
       series = {Astronomical Society of the Pacific Conference Series},
       volume = {534},
        month = jul,
        pages = {275},
          doi = {10.48550/arXiv.2203.10066},
archivePrefix = {arXiv},
       eprint = {2203.10066},
 primaryClass = {astro-ph.SR},
       adsurl = {https://ui.adsabs.harvard.edu/abs/2023ASPC..534..275O}
}

@ARTICLE{2025A&A...695A..61M,
       author = {{Merc}, Jaroslav and {Boffin}, Henri M.~J.},
        title = "{Revisiting symbiotic binaries with interferometry: I. The PIONIER archival collection}",
      journal = {\aap},
         year = 2025,
        month = mar,
       volume = {695},
          eid = {A61},
        pages = {A61},
          doi = {10.1051/0004-6361/202553789},
archivePrefix = {arXiv},
       eprint = {2502.04089},
 primaryClass = {astro-ph.SR},
       adsurl = {https://ui.adsabs.harvard.edu/abs/2025A&A...695A..61M}
}

@ARTICLE{2020MNRAS.495..321P,
       author = {{Penoyre}, Zephyr and {Belokurov}, Vasily and {Wyn Evans}, N. and {Everall}, A. and {Koposov}, S.~E.},
        title = "{Binary deviations from single object astrometry}",
      journal = {\mnras},
         year = 2020,
        month = jun,
       volume = {495},
       number = {1},
        pages = {321-337},
          doi = {10.1093/mnras/staa1148},
archivePrefix = {arXiv},
       eprint = {2003.05456},
 primaryClass = {astro-ph.GA},
       adsurl = {https://ui.adsabs.harvard.edu/abs/2020MNRAS.495..321P}
}

@ARTICLE{2021AJ....161..147B,
       author = {{Bailer-Jones}, C.~A.~L. and {Rybizki}, J. and {Fouesneau}, M. and {Demleitner}, M. and {Andrae}, R.},
        title = "{Estimating Distances from Parallaxes. V. Geometric and Photogeometric Distances to 1.47 Billion Stars in Gaia Early Data Release 3}",
      journal = {\aj},
         year = 2021,
        month = mar,
       volume = {161},
       number = {3},
          eid = {147},
        pages = {147},
          doi = {10.3847/1538-3881/abd806},
archivePrefix = {arXiv},
       eprint = {2012.05220},
 primaryClass = {astro-ph.SR},
       adsurl = {https://ui.adsabs.harvard.edu/abs/2021AJ....161..147B}
}

@PHDTHESIS{2022PhDT........34M,
       author = {{Merc}, Jaroslav},
        title = "{Multi-frequency research of symbiotic binaries}",
       school = {Charles University in Prague / P. J. {\v{S}}af{\'a}rik University in
        Ko{\v{s}}ice, Czech Republic / Slovakia},
         year = 2022,
        month = jan,
       adsurl = {https://ui.adsabs.harvard.edu/abs/2022PhDT........34M}
}

@ARTICLE{2025A&A...699A.117M,
       author = {{Merc}, Jaroslav and {Guerrero}, Mart{\'\i}n A. and {Toal{\'a}}, Jes{\'u}s A. and {Ortiz}, Roberto},
        title = "{CGCS 6306, another X-ray-emitting asymptotic giant branch star confirmed to be a symbiotic binary}",
      journal = {\aap},
         year = 2025,
        month = jul,
       volume = {699},
          eid = {A117},
        pages = {A117},
          doi = {10.1051/0004-6361/202554671},
archivePrefix = {arXiv},
       eprint = {2505.24660},
 primaryClass = {astro-ph.SR},
       adsurl = {https://ui.adsabs.harvard.edu/abs/2025A&A...699A.117M}
}

@article{Merc+SALT_PaperII,
    author = {{Merc}, J. and {Miko\l{}ajewska}, J. and {I\l{}kiewicz}, K. and {Monard}, B.},
    title = "{Identification of new Galactic symbiotic stars with SALT - II. New discoveries and characterization of the sample}",
    journal = {submitted to MNRAS},
    year = 2025
}

@article{Merc+Gaia2,
    author = {{Merc}, J. and {Mulato}, L. and {Charbonnel}, St{\'e}phane and {Garde}, Olivier and {Le D\^{u}}, Pascal and {Petit}, Thomas},
    title = "{Gaia DR3 candidates}",
    journal = {in preparation},
    year = 2026
}

@article{Merc+Gaia3,
    author = {{Merc}, J. and {Mulato}, L. and {Charbonnel}, St{\'e}phane and {Garde}, Olivier and {Le D\^{u}}, Pascal and {Petit}, Thomas},
    title = "{New Gaia DR3 symbiotics}",
    journal = {in preparation},
    year = 2026
}

@article{Mulato+PNe,
    author = {{Mulato}, L. and {Merc}, J. and {Charbonnel}, St{\'e}phane and {Garde}, Olivier and {Le D\^{u}}, Pascal and {Petit}, Thomas},
    title = "{Analysis of the \textit{Gaia} DR3 planetary nebula candidates and the possible symbiotic stars among them}",
    journal = {submitted to A\&A},
    year = 2026
}

@ARTICLE{2026MNRAS.545S2146M,
       author = {{Merc}, J. and {Miko{\l}ajewska}, J. and {I{\l}kiewicz}, K. and {Monard}, B. and {Udalski}, A.},
        title = "{Identification of new Galactic symbiotic stars with SALT─II. New discoveries and characterization of the sample}",
      journal = {\mnras},
         year = 2026,
        month = feb,
       volume = {545},
       number = {4},
          eid = {staf2146},
        pages = {staf2146},
          doi = {10.1093/mnras/staf2146},
archivePrefix = {arXiv},
       eprint = {2512.01744},
 primaryClass = {astro-ph.SR},
       adsurl = {https://ui.adsabs.harvard.edu/abs/2026MNRAS.545S2146M}
}

@ARTICLE{2023A&A...674A..28F,
       author = {{Fouesneau}, M. and {Fr{\'e}mat}, Y. and {Andrae}, R. and {Korn}, A.~J. and {Soubiran}, C. and {Kordopatis}, G. and {Vallenari}, A. and {Heiter}, U. and {Creevey}, O.~L. and {Sarro}, L.~M. and {de Laverny}, P. and {Lanzafame}, A.~C. and {Lobel}, A. and {Sordo}, R. and {Rybizki}, J. and {Slezak}, I. and {{\'A}lvarez}, M.~A. and {Drimmel}, R. and {Garabato}, D. and {Delchambre}, L. and {Bailer-Jones}, C.~A.~L. and {Hatzidimitriou}, D. and {Lorca}, A. and {Le Fustec}, Y. and {Pailler}, F. and {Mary}, N. and {Robin}, C. and {Utrilla}, E. and {Abreu Aramburu}, A. and {Bakker}, J. and {Bellas-Velidis}, I. and {Bijaoui}, A. and {Blomme}, R. and {Bouret}, J. -C. and {Brouillet}, N. and {Brugaletta}, E. and {Burlacu}, A. and {Carballo}, R. and {Casamiquela}, L. and {Chaoul}, L. and {Chiavassa}, A. and {Contursi}, G. and {Cooper}, W.~J. and {Dafonte}, C. and {Demouchy}, C. and {Dharmawardena}, T.~E. and {Garc{\'\i}a-Lario}, P. and {Garc{\'\i}a-Torres}, M. and {Gomez}, A. and {Gonz{\'a}lez-Santamar{\'\i}a}, I. and {Jean-Antoine Piccolo}, A. and {Kontizas}, M. and {Lebreton}, Y. and {Licata}, E.~L. and {Lindstr{\o}m}, H.~E.~P. and {Livanou}, E. and {Magdaleno Romeo}, A. and {Manteiga}, M. and {Marocco}, F. and {Martayan}, C. and {Marshall}, D.~J. and {Nicolas}, C. and {Ordenovic}, C. and {Palicio}, P.~A. and {Pallas-Quintela}, L. and {Pichon}, B. and {Poggio}, E. and {Recio-Blanco}, A. and {Riclet}, F. and {Santove{\~n}a}, R. and {Schultheis}, M.~S. and {Segol}, M. and {Silvelo}, A. and {Smart}, R.~L. and {S{\"u}veges}, M. and {Th{\'e}venin}, F. and {Torralba Elipe}, G. and {Ulla}, A. and {van Dillen}, E. and {Zhao}, H. and {Zorec}, J.},
        title = "{Gaia Data Release 3. Apsis. II. Stellar parameters}",
      journal = {\aap},
         year = 2023,
        month = jun,
       volume = {674},
          eid = {A28},
        pages = {A28},
          doi = {10.1051/0004-6361/202243919},
archivePrefix = {arXiv},
       eprint = {2206.05992},
 primaryClass = {astro-ph.SR},
       adsurl = {https://ui.adsabs.harvard.edu/abs/2023A&A...674A..28F}
}

@ARTICLE{2023A&A...674A..26C,
       author = {{Creevey}, O.~L. and {Sordo}, R. and {Pailler}, F. and {Fr{\'e}mat}, Y. and {Heiter}, U. and {Th{\'e}venin}, F. and {Andrae}, R. and {Fouesneau}, M. and {Lobel}, A. and {Bailer-Jones}, C.~A.~L. and {Garabato}, D. and {Bellas-Velidis}, I. and {Brugaletta}, E. and {Lorca}, A. and {Ordenovic}, C. and {Palicio}, P.~A. and {Sarro}, L.~M. and {Delchambre}, L. and {Drimmel}, R. and {Rybizki}, J. and {Torralba Elipe}, G. and {Korn}, A.~J. and {Recio-Blanco}, A. and {Schultheis}, M.~S. and {De Angeli}, F. and {Montegriffo}, P. and {Abreu Aramburu}, A. and {Accart}, S. and {{\'A}lvarez}, M.~A. and {Bakker}, J. and {Brouillet}, N. and {Burlacu}, A. and {Carballo}, R. and {Casamiquela}, L. and {Chiavassa}, A. and {Contursi}, G. and {Cooper}, W.~J. and {Dafonte}, C. and {Dapergolas}, A. and {de Laverny}, P. and {Dharmawardena}, T.~E. and {Edvardsson}, B. and {Le Fustec}, Y. and {Garc{\'\i}a-Lario}, P. and {Garc{\'\i}a-Torres}, M. and {Gomez}, A. and {Gonz{\'a}lez-Santamar{\'\i}a}, I. and {Hatzidimitriou}, D. and {Jean-Antoine Piccolo}, A. and {Kontiza}, M. and {Kordopatis}, G. and {Lanzafame}, A.~C. and {Lebreton}, Y. and {Licata}, E.~L. and {Lindstr{\o}m}, H.~E.~P. and {Livanou}, E. and {Magdaleno Romeo}, A. and {Manteiga}, M. and {Marocco}, F. and {Marshall}, D.~J. and {Mary}, N. and {Nicolas}, C. and {Pallas-Quintela}, L. and {Panem}, C. and {Pichon}, B. and {Poggio}, E. and {Riclet}, F. and {Robin}, C. and {Santove{\~n}a}, R. and {Silvelo}, A. and {Slezak}, I. and {Smart}, R.~L. and {Soubiran}, C. and {S{\"u}veges}, M. and {Ulla}, A. and {Utrilla}, E. and {Vallenari}, A. and {Zhao}, H. and {Zorec}, J. and {Barrado}, D. and {Bijaoui}, A. and {Bouret}, J. -C. and {Blomme}, R. and {Brott}, I. and {Cassisi}, S. and {Kochukhov}, O. and {Martayan}, C. and {Shulyak}, D. and {Silvester}, J.},
        title = "{Gaia Data Release 3. Astrophysical parameters inference system (Apsis). I. Methods and content overview}",
      journal = {\aap},
         year = 2023,
        month = jun,
       volume = {674},
          eid = {A26},
        pages = {A26},
          doi = {10.1051/0004-6361/202243688},
archivePrefix = {arXiv},
       eprint = {2206.05864},
 primaryClass = {astro-ph.GA},
       adsurl = {https://ui.adsabs.harvard.edu/abs/2023A&A...674A..26C}
}

@ARTICLE{2023A&A...674A..27A,
       author = {{Andrae}, R. and {Fouesneau}, M. and {Sordo}, R. and {Bailer-Jones}, C.~A.~L. and {Dharmawardena}, T.~E. and {Rybizki}, J. and {De Angeli}, F. and {Lindstr{\o}m}, H.~E.~P. and {Marshall}, D.~J. and {Drimmel}, R. and {Korn}, A.~J. and {Soubiran}, C. and {Brouillet}, N. and {Casamiquela}, L. and {Rix}, H. -W. and {Abreu Aramburu}, A. and {{\'A}lvarez}, M.~A. and {Bakker}, J. and {Bellas-Velidis}, I. and {Bijaoui}, A. and {Brugaletta}, E. and {Burlacu}, A. and {Carballo}, R. and {Chaoul}, L. and {Chiavassa}, A. and {Contursi}, G. and {Cooper}, W.~J. and {Creevey}, O.~L. and {Dafonte}, C. and {Dapergolas}, A. and {de Laverny}, P. and {Delchambre}, L. and {Demouchy}, C. and {Edvardsson}, B. and {Fr{\'e}mat}, Y. and {Garabato}, D. and {Garc{\'\i}a-Lario}, P. and {Garc{\'\i}a-Torres}, M. and {Gavel}, A. and {Gomez}, A. and {Gonz{\'a}lez-Santamar{\'\i}a}, I. and {Hatzidimitriou}, D. and {Heiter}, U. and {Jean-Antoine Piccolo}, A. and {Kontizas}, M. and {Kordopatis}, G. and {Lanzafame}, A.~C. and {Lebreton}, Y. and {Licata}, E.~L. and {Livanou}, E. and {Lobel}, A. and {Lorca}, A. and {Magdaleno Romeo}, A. and {Manteiga}, M. and {Marocco}, F. and {Mary}, N. and {Nicolas}, C. and {Ordenovic}, C. and {Pailler}, F. and {Palicio}, P.~A. and {Pallas-Quintela}, L. and {Panem}, C. and {Pichon}, B. and {Poggio}, E. and {Recio-Blanco}, A. and {Riclet}, F. and {Robin}, C. and {Santove{\~n}a}, R. and {Sarro}, L.~M. and {Schultheis}, M.~S. and {Segol}, M. and {Silvelo}, A. and {Slezak}, I. and {Smart}, R.~L. and {S{\"u}veges}, M. and {Th{\'e}venin}, F. and {Torralba Elipe}, G. and {Ulla}, A. and {Utrilla}, E. and {Vallenari}, A. and {van Dillen}, E. and {Zhao}, H. and {Zorec}, J.},
        title = "{Gaia Data Release 3. Analysis of the Gaia BP/RP spectra using the General Stellar Parameterizer from Photometry}",
      journal = {\aap},
         year = 2023,
        month = jun,
       volume = {674},
          eid = {A27},
        pages = {A27},
          doi = {10.1051/0004-6361/202243462},
archivePrefix = {arXiv},
       eprint = {2206.06138},
 primaryClass = {astro-ph.SR},
       adsurl = {https://ui.adsabs.harvard.edu/abs/2023A&A...674A..27A}
}

@ARTICLE{2023A&A...677A..37C,
       author = {{Castro-Ginard}, Alfred and {Brown}, Anthony G.~A. and {Kostrzewa-Rutkowska}, Zuzanna and {Cantat-Gaudin}, Tristan and {Drimmel}, Ronald and {Oh}, Semyeong and {Belokurov}, Vasily and {Casey}, Andrew R. and {Fouesneau}, Morgan and {Khanna}, Shourya and {Price-Whelan}, Adrian M. and {Rix}, Hans-Walter},
        title = "{Estimating the selection function of Gaia DR3 subsamples}",
      journal = {\aap},
         year = 2023,
        month = sep,
       volume = {677},
          eid = {A37},
        pages = {A37},
          doi = {10.1051/0004-6361/202346547},
archivePrefix = {arXiv},
       eprint = {2303.17738},
 primaryClass = {astro-ph.GA},
       adsurl = {https://ui.adsabs.harvard.edu/abs/2023A&A...677A..37C}
}

@MISC{2012ivoa.rept.1015R,
       author = {{Rodrigo}, Carlos and {Solano}, Enrique and {Bayo}, Amelia},
        title = "{SVO Filter Profile Service Version 1.0}",
 howpublished = {IVOA Working Draft 15 October 2012},
         year = 2012,
        month = oct,
        pages = {1015},
          doi = {10.5479/ADS/bib/2012ivoa.rept.1015R},
       adsurl = {https://ui.adsabs.harvard.edu/abs/2012ivoa.rept.1015R}
}

@INPROCEEDINGS{2020sea..confE.182R,
       author = {{Rodrigo}, C. and {Solano}, E.},
        title = "{The SVO Filter Profile Service}",
    booktitle = {XIV.0 Scientific Meeting (virtual) of the Spanish Astronomical Society},
         year = 2020,
        month = jul,
          eid = {182},
        pages = {182},
       adsurl = {https://ui.adsabs.harvard.edu/abs/2020sea..confE.182R}
}


\section*{Supporting information}
Supplementary data are available online.\\

\noindent \textbf{Table 1.} List of confirmed symbiotic stars analysed in this work, together with their corresponding \textit{Gaia}~DR3 identifiers.

\FloatBarrier
\appendix
\begin{figure}
\section{Additional figures}\label{app:fig}
\centering
\includegraphics[width=\columnwidth]{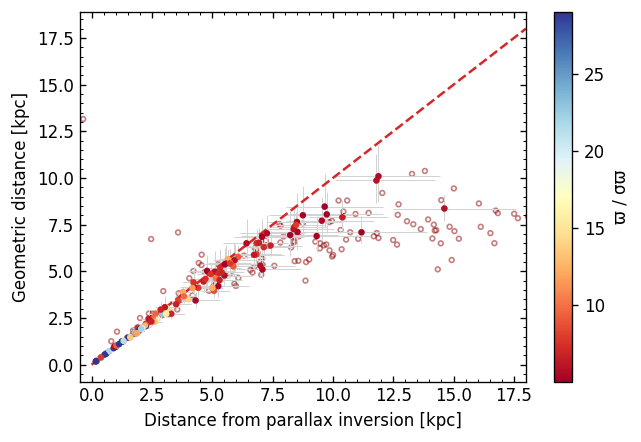}

\caption{Comparison of distances derived from direct parallax inversion with geometric distances from \citet{2021AJ....161..147B}. The red dashed line indicates the 1:1 relation. Data points are color-coded by parallax signal-to-noise ratio, with objects having S/N < 5 shown as empty circles.}
\label{fig:dist}
\end{figure}

\begin{figure}
\centering
\includegraphics[width=\columnwidth]{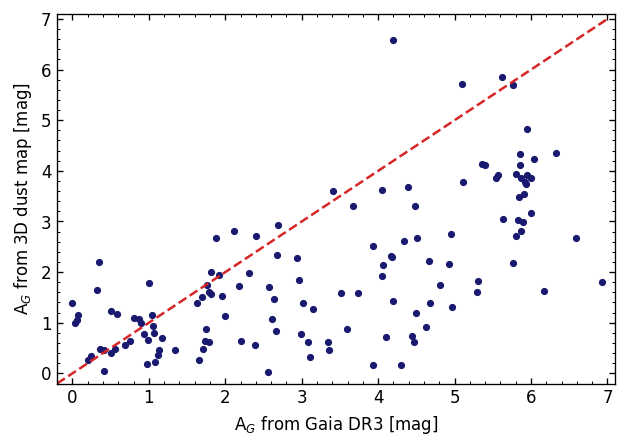}
\includegraphics[width=\columnwidth]{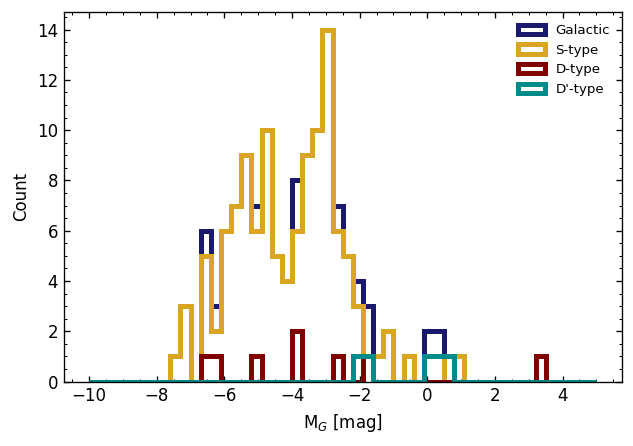}

\caption{\textbf{Upper panel:} Comparison of $G$-band extinction from \textit{Gaia} DR3 ({\tt astrophysical\_parameters} table, derived using the GSP-Phot Aeneas best-library solution from $BP/RP$ spectra) with values from 3D dust maps (see text for details). The red
dashed line indicates the 1:1 relation.
\textbf{Lower panel:} Distribution of $M_G$ calculated using \textit{Gaia}-based extinction, analogous to Fig.~\ref{fig:mags}E. Extinction values from \textit{Gaia} introduce significantly larger scatter.}
\label{fig:extinc}
\end{figure}

\begin{figure}
\centering
\includegraphics[width=\columnwidth]{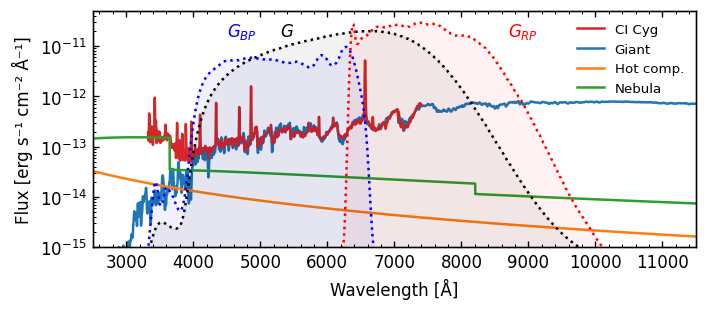}

\caption{\textit{Gaia} DR3 filter response curves, shown together with the observed spectrum of CI Cyg and a model consisting of a red giant, nebular continuum, and a hot companion represented by a blackbody. The observed spectrum is taken from the ARAS database \citep[][]{2019CoSka..49..217T}, the red giant spectrum corresponds to an M4 star \citep[][]{1998PASP..110..863P}, the nebular continuum was computed with the {\tt ChiantiPy} package \citep[][]{1997A&AS..125..149D,2021ApJ...909...38D}, and the filter response curves were downloaded from the Spanish Virtual Observatory (SVO) Filter Profile Service \citep[][]{2020sea..confE.182R,2012ivoa.rept.1015R,2024A&A...689A..93R}. }
\label{fig:dr3_filters}
\end{figure}

\begin{figure}
\centering
\includegraphics[width=\columnwidth]{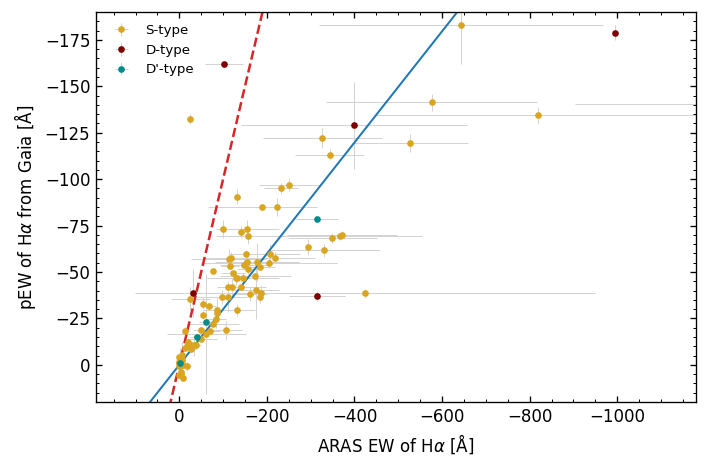}

\caption{Comparison of \textit{Gaia} DR3 pEW of H$\alpha$ with values obtained from ARAS spectra \citep[][]{2019CoSka..49..217T}. The red dashed line marks the 1:1 relation, while the solid blue line shows a linear fit to the data (restricted to ARAS EW $> -400$ \AA{}) and forced to pass through (0, 0).}
\label{fig:halpha_ARAS}
\end{figure}

\begin{figure}
\section{First orbits of StHA 154 and StHA 151}\label{sec:orbits}
\centering
\includegraphics[width=\columnwidth]{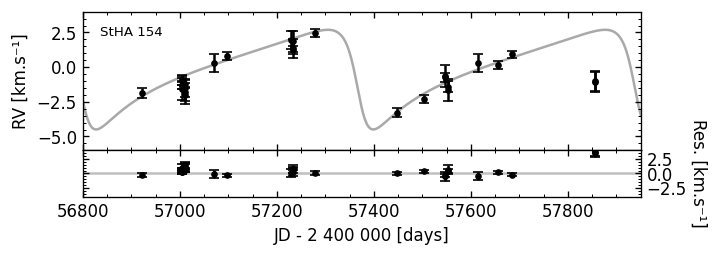}
\includegraphics[width=\columnwidth]{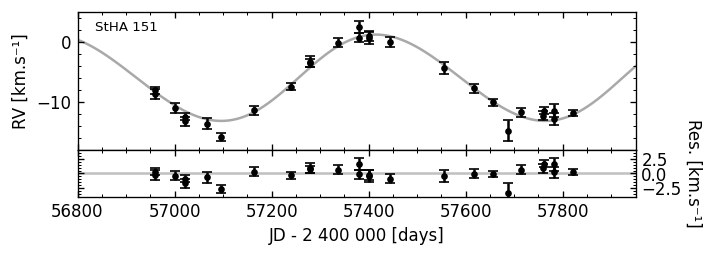}
\caption{Orbits of StHA 154 and StHA 151.}
\label{fig:orbits}
\end{figure}

\begin{table}
\centering
\small
\caption{Orbital elements of StHA 154 and StHA 151.\label{tab:orbit}}%
\begin{tabular}{lll}
\hline
\noalign{\smallskip}
 & StHA 154 & StHA 151  \\
 \hline
  \noalign{\smallskip}
$P$ {[}d{]} & 571.4 $\pm$ 13.5 & 667.9 $\pm$ 9.5  \\
$e$ & 0.584 $\pm$ 0.186 & 0.037 $\pm$ 0.031 \\
$T_0$ 2 4.. {[}d{]} & 57370 $\pm$ 119 & 57285 $\pm$ 33  \\
$K$ {[}km/s{]} & 3.60 $\pm$ 1.69 & 7.17 $\pm$ 0.28  \\
$\omega$ {[}\textdegree{]} & 110.33 $\pm$ 12.38 & -89.59 $\pm$ 4.23  \\
$\gamma [km/s]$ & -0.17 $\pm$ 0.59 & -5.96 $\pm$ 0.04\\
$f$(m) {[}M$_\odot${]} & 0.001479 $\pm$ 0.002207  &  0.025456 $\pm$ 0.003006 \\
$\chi^2$ & 4.35 & 2.28 \\
\noalign{\smallskip}
\hline
\end{tabular}
\end{table}
\FloatBarrier

\bsp	
\label{lastpage}
\end{document}